\documentclass[final,5p,times,twocolumn]{elsarticle}
\usepackage{amssymb,amsmath,graphicx,xcolor,slashed}
\usepackage[colorlinks=true,linkcolor=blue,citecolor=blue,urlcolor=blue]{hyperref}
\usepackage{cleveref}
\usepackage{physics}
\usepackage{subfigure}
\usepackage{multirow}
\usepackage{comment}
\usepackage{longtable}
\usepackage{pdflscape}
\usepackage{diagbox}
\usepackage{soul,xcolor}

\newcommand{\LambdaQCD}{\Lambda_\text{QCD}}
\newcommand{\N}{NLO}
\newcommand{\NL}{NLO+LRR}

\newcommand{\NLR}{NLR}
\newcommand{\diff}[2]{\operatorname{d}{\hspace{-0.15em}}^{#1}{#2}\hspace{0.15em}}

\journal{Physics Letters B}

\begin{document}
\begin{frontmatter}

\title{Nucleon Helicity Parton Distribution Function in the Continuum Limit with Self-Renormalization}

\author[first]{Jack Holligan}
\affiliation[first]{organization={Michigan State University},
            addressline={Department of Physics and Astronomy, Michigan State University}, 
            city={East Lansing},
            postcode={48824}, 
            state={MI},
            country={USA}}

\author[first]{Huey-Wen Lin}

\begin{abstract}
We present the first lattice calculation of the nucleon isovector helicity parton distribution function (PDF) in the framework of large-momentum effective theory (LaMET) that uses the hybrid scheme with self-renormalization.
We use ensembles generated by the MILC collaboration at lattice spacings $a=\{0.1207,0.0888,0.0582\}$~fm, with $N_f=2+1+1$ flavors of highly improved staggered quarks at sea pion mass of $M_{\pi}\approx 315$~MeV.
We use clover-improved action for our valence quarks with nucleon boost momentum $P_z\approx 1.75$~GeV and high-statistics measurements for the LaMET matrix elements. We perform an extrapolation to the continuum limit and improve the handling of systematic errors using renormalization-group resummation (RGR) and leading-renormalon resummation (LRR).
Our final nucleon helicity PDF is renormalized in the $\overline{\text{MS}}$ scheme at energy scale $\mu=2.0$~GeV.
We compare our results with and without the two systematic improvements of RGR and LRR at each lattice spacing as well as the continuum limit, and we see that the application of RGR and LRR greatly reduces the systematic errors across the whole $x$ range.
Our continuum results with both RGR and LRR show a small positive antiquark region for the nucleon helicity PDF as well as a change of as much as a factor of two in the central values compared to results with neither RGR or LRR.
By contrast, the application of RGR and LRR only changes the central values by about 5\% in the quark region.
We compare our lattice results with the global fits by the JAM, NNPDF and DSSV collaborations, and we observe some tension between our results.
\end{abstract}


\end{frontmatter}

\section{Introduction}\label{sec:Introduction}

Parton distribution functions (PDFs) describe non perturbatively the probability distribution of specific longitudinal momentum fractions, $x$, of a hadron's constituent quarks and gluons. 
Among them, helicity PDFs 
provide information on the difference between the parton having its spin aligned and opposite to the hadron's spin. 
Experimentally, great progress has been made in the study of nucleon helicity PDFs 
through semi-inclusive deep-inelastic scattering and high-energy muon scattering experiments, such as by the HERMES~\cite{HERMES:2004zsh} and COMPASS~\cite{COMPASS:2009kiy,COMPASS:2010hwr} collaborations at DESY and CERN, respectively. 
Other experiments, such as the STAR~\cite{STAR:2010xwx,STAR:2012hth,STAR:2014afm,STAR:2014wox} and PHENIX~\cite{PHENIX:2010aru,PHENIX:2010rkr,PHENIX:2015ade} collaborations which study proton-proton collisions at the Relativistic Heavy Ion Collider (RHIC) are able to probe the antiquark contribution to the helicity PDF. 
Future experiments such as those that have been proposed at the Electron Ion Collider (EIC)~\cite{AbdulKhalek:2021gbh,Achenbach:2023pba,Abir:2023fpo,AbdulKhalek:2022hcn,Burkert:2022hjz} in the United States as well as the Electron Ion Collider in China (EicC)~\cite{Anderle:2021dpv} and the AMBER experiment at CERN~\cite{Quintans:2022utc} will probe the spin structure of the nucleon with even greater accuracy and help refine the global fits of the helicity 
PDFs. 

Progress in global fits  has been made in extracting helicity PDFs from experimental data by the DSSV~\cite{deFlorian:2009vb}, NNPDF~\cite{Nocera:2014gqa} and JAM~\cite{Ethier:2017zbq} collaborations. 
DSSV~\cite{deFlorian:2009vb} showed that there is a positive antiquark region for the proton helicity PDF ($\Delta\overline{u}(x)-\Delta\overline{d}(x)>0$) as well as that quarks and antiquarks only contribute up to one third of the proton's spin. 
NNPDF~\cite{Nocera:2014gqa} extended the proton helicity PDF calculation to the small-$x$ region for polarized gluons as well as performed a calculation of the individual unpolarized light quark and antiquark PDFs. 
JAM~\cite{Ethier:2017zbq} determined that the contribution to the proton's spin from strange quarks is compatible with zero, which provided more information about the total spin content of the proton.
However, assumptions often have to be made in the global analysis of helicity PDFs, such as the antiquark region being zero, the functional form of the PDF, or exact SU(3) flavor symmetry. 
Both the DSSV~\cite{deFlorian:2009vb} and NNPDF~\cite{Nocera:2014gqa} collaborations made the assumption $\Delta s=\Delta\overline{s}$ in their calculations. 
JAM relaxed this assumption by including semi-inclusive deep inelastic scattering (SIDIS) data~\cite{Ethier:2017zbq} but found that any deviation from zero was unable to rise above background noise. 
While global fitting has the advantage of using experimental data from real events, it does not perform a direct PDF calculation from the first principles of quantum chromodynamics (QCD) and is limited by the quality of experimental data that is available. 

A direct calculation of PDFs became possible with the advent of large-momentum effective theory (LaMET) in 2013~\cite{Ji:2013dva,Ji:2014gla,Ji:2020ect}, which allowed parton physics to be studied on the Euclidean lattice. 
By studying the behavior of spatially separated correlators boosted to large momentum, parton physics can then be recovered through perturbative matching. 
The bare matrix elements extracted from the lattice are $h^B(z,a)=\left\langle P_z\left|\bar\psi\left(-\frac{z}{2}\right)\Gamma W\left(-\frac{z}{2},\frac{z}{2}\right)\psi\left(\frac{z}{2}\right)\right|P_z\right\rangle$, where $\ket{P_z}$ is a hadron state at boost momentum $P_z$ in the $z$ direction, $W(-\frac{z}{2},\frac{z}{2})=\exp[ig\int_{-z/2}^{z/2}\diff{}{z'}A_z(z')]$ is the Wilson line connecting the two spacetime coordinates $(0,0,-z/2,0)$ and $(0,0,z/2,0)$, $\psi$ is the fermion field, and $\Gamma$ is a Dirac structure which for helicity matrix elements is $\Gamma=\gamma_z\gamma_5$.
To extract the ground-state matrix element, we use a two-state fit on the two-point correlators and a two-sim fit on the three-point correlators, following, for example, Ref.~\cite{Fan:2022kcb}. The extraction of the matrix element can be visualized by plotting the ratio of the three-point and two-point correlation functions at multiple values of source-sink separation $t_\text{sep}$. The ratio is observed to approach the computed ground-state matrix element as $t_\text{sep}$ increases. In addition, we vary the minimum and maximum values of $t_\text{sep}$ (denoted by $t_\text{sep}^\text{min}$ and $t_\text{sep}^\text{max}$), again following the example of Ref.~\cite{Fan:2022kcb}. We find that the ground-state matrix elements are compatible for these different values. This shows that excited-state contamination is under control and not a significant source of systematic error.
The LaMET method has been used for a numerical determination of the nucleon helicity PDF in Refs.~\cite{Chen:2016utp,Lin:2019ocg,Lin:2018pvv,Alexandrou:2020qtt,Alexandrou:2016jqi,Alexandrou:2017huk,Lin:2017ani,Alexandrou:2018pbm,Fan:2020nzz}.  
The very first helicity PDF calculation was done in 2016~\cite{Chen:2016utp} using 310 MeV pion mass at lattice spacing $a=0.12$ fm 
with boost momenta $P_z=\{0.43,\,0.86,\,1.29\}$ GeV. Mass correction and one-loop matching was used to better align the behavior of the quasi-PDF with that of the lightcone. 
This was the first time the antiquark section had been studied and showed an asymmetry $\overline{u}(x)>\overline{d}(x)$. 
Soon after, there were many followup LaMET calculations by different collaborations with different fermion actions and even physical pion masses~\cite{Lin:2018pvv,Alexandrou:2020qtt,Alexandrou:2016jqi,Alexandrou:2017huk,Lin:2017ani,Alexandrou:2018pbm,Fan:2020nzz,HadStruc:2022nay}.
Most LaMET-method helicity PDFs have been renormalized nonperturbatively only in the RI/MOM and similar renormalization schemes~\cite{Lin:2018pvv,Alexandrou:2020qtt,Alexandrou:2017huk,Lin:2017ani,Alexandrou:2018pbm,Fan:2020nzz}.  
The latest hybrid- or self-renormalization schemes~\cite{Ji:2020brr,LatticePartonCollaborationLPC:2021xdx} have not been applied to the helicity PDF yet.

In this work, we make the first calculation of the isovector nucleon helicity PDF using matrix elements renormalized in the hybrid scheme with self-renormalization (HSR). The hybrid scheme was introduced in Ref.~\cite{Ji:2020brr} and the self-renormalization in Ref.~\cite{LatticePartonCollaborationLPC:2021xdx}.
We use this method to address two sources of divergence in the matrix elements: the linear divergence which arises from the self-energy in the Wilson line of the bare matrix elements and the renormalon divergence which arises from the asymptotic nature of the perturbation series.
This is done by fitting the matrix elements to a functional form derived from perturbative QCD which has the added advantage of accounting for discretization effects.
It extracts the renormalization factor and the remaining non-perturbative physics directly from the matrix elements.
It also reduces the dependence on lattice spacing $a$ during the renormalization process, which allows for a reliable extrapolation to the continuum limit.
These two sources of divergence can also be addressed using the hybrid scheme introduced in Ref.~\cite{Ji:2020brr}.
However, this method does not account directly for discretization effects, and the removal of the linear divergence in this scheme is a very delicate exercise and a significant source of systematic error, as shown in Ref.~\cite{LatticePartonCollaborationLPC:2021xdx}.
The HSR method has been used in the calculation of nucleon transversity PDFs~\cite{LatticeParton:2022xsd} as well as the calculation of pion and kaon distribution amplitudes~\cite{Holligan:2023rex,LatticeParton:2022zqc}.
In both cases, the HSR process greatly reduces the dependence on lattice spacing compared with pure RI/MOM scheme.

We can further improve our calculation by augmenting both the renormalization scheme and the lightcone matching with the methods of renormalization-group resummation (RGR)~\cite{Holligan:2023rex,Su:2022fiu,Zhang:2023bxs} and leading-renormalon resummation (LRR)~\cite{Zhang:2023bxs}.
RGR involves resumming the large logarithms that arise from the difference in renormalization scale and the intrinsic physical scale.
The method involves setting the renormalization scale such that the logarithms vanish and then evolving to the desired energy scale using the renormalization-group equation (RGE).
The perturbation series also contains a renormalon divergence~\cite{Zichichi:1979gj}, which is enhanced by the use of RGR on its own.
We account for this effect by including LRR in the calculations which resums the series to account for the renormalon divergence.
We present the first application of these methods to the nucleon helicity PDF.

\section{Self-Renormalization in Hybrid Scheme}
\label{sec:SelfRenormalization}

\begin{table*}[htbp!]
\centering
\begin{tabular}{|c|c|c|c|c|c|c|c|c|c|c|}
\hline
Ensemble ID & $a$ (fm) & $N_s^3 \times N_t$& $M_\pi^\text{val}$ (MeV)  & $t_\text{sep}/a$ & $P_z$ (GeV)   & $N_\text{meas}$  \\
\hline\hline
a12m310  & 0.1207(11) & $24^3\times 64$  & 310(3)  & $\{6,7,8,9\}$      &  $1.71$  &  $\{50904, 50904, 101808, 203616\}$   \\
\hline
a09m310  & 0.0888(08)  & $32^3\times 96$ & 313(1)  &  $\{8,9,10,12\}$   &  $1.75$  &  $\{109616,109616,219232,328848\}$ \\
\hline
a06m310  & 0.0582(4)  & $48^3\times 96$  & 320(2) & $\{12,14,16,18\}$  &  $1.78$  &  $\{179520,269280,359040,538560\}$ \\
\hline
\end{tabular}
\caption{\label{tab:params}
Ensemble information and parameters used in this work.
$N_\text{meas}$ is the total number of measurements of the three-point correlators for different values of $t_\text{sep}$.
$L$ indicates the spatial length which is $aN_s$ (in fm).
}
\end{table*}

In this work, we use clover lattice fermion action for the valence quarks on top of 2+1+1 flavors (degenerate up and down quarks plus strange and charm quarks at their physical masses in the QCD vacuum) of hypercubic (HYP)-smeared~\cite{Hasenfratz:2001hp} HISQ \cite{MILC:2012znn,Follana:2006rc}, generated by MILC Collaboration.
The lattice parameters include lattice spacings $a \in [0.06,0.12]$~fm, pion mass $M_\pi \approx 310$~MeV and $M_\pi L \approx 4.5$.
The quark masses for the clover action are tuned to reproduce the lightest sea HISQ pseudoscalar meson masses, and the clover parameters are set to the tree-level tadpole-improved values.
We carefully monitor for signatures of exceptional configurations due to the non-unitary lattice-QCD formulation of mixed-action approach and found exceptional configurations to be absent for these three MILC ensembles~\cite{MILC:2012znn}.
This mixed-action setup is the same as the one used in works done by PNDME Collaboration in many studies of nucleon structure~\cite{Gupta:2018qil,Bhattacharya:2015wna,Bhattacharya:2015esa,Bhattacharya:2013ehc}.

On the lattice, we calculate the time-independent, nonlocal matrix elements stated in Sec.~\ref{sec:Introduction}.
We use Gaussian momentum smearing~\cite{Bali:2016lva} for the quark field.
Such a momentum source is designed to increase the overlap with nucleons of the desired boost momentum, and we are able to reach higher boost momentum for the nucleon states.
We use multigrid algorithm~\cite{Babich:2010qb,Osborn:2010mb} in the Chroma software package~\cite{Edwards:2004sx} to speed up the inversion of the quark propagator for the clover fermions.
To make sure excited-state contamination is under control, we measure at least four nucleon three-point source-sink separations, and we perform simultaneously two-state extraction of ground-state nucleon matrix elements.
Details of our calculation parameters can be found in Table~\ref{tab:params}.

\begin{figure*}[ht]
  \centering
  \subfigure{\includegraphics[width=0.3\linewidth]{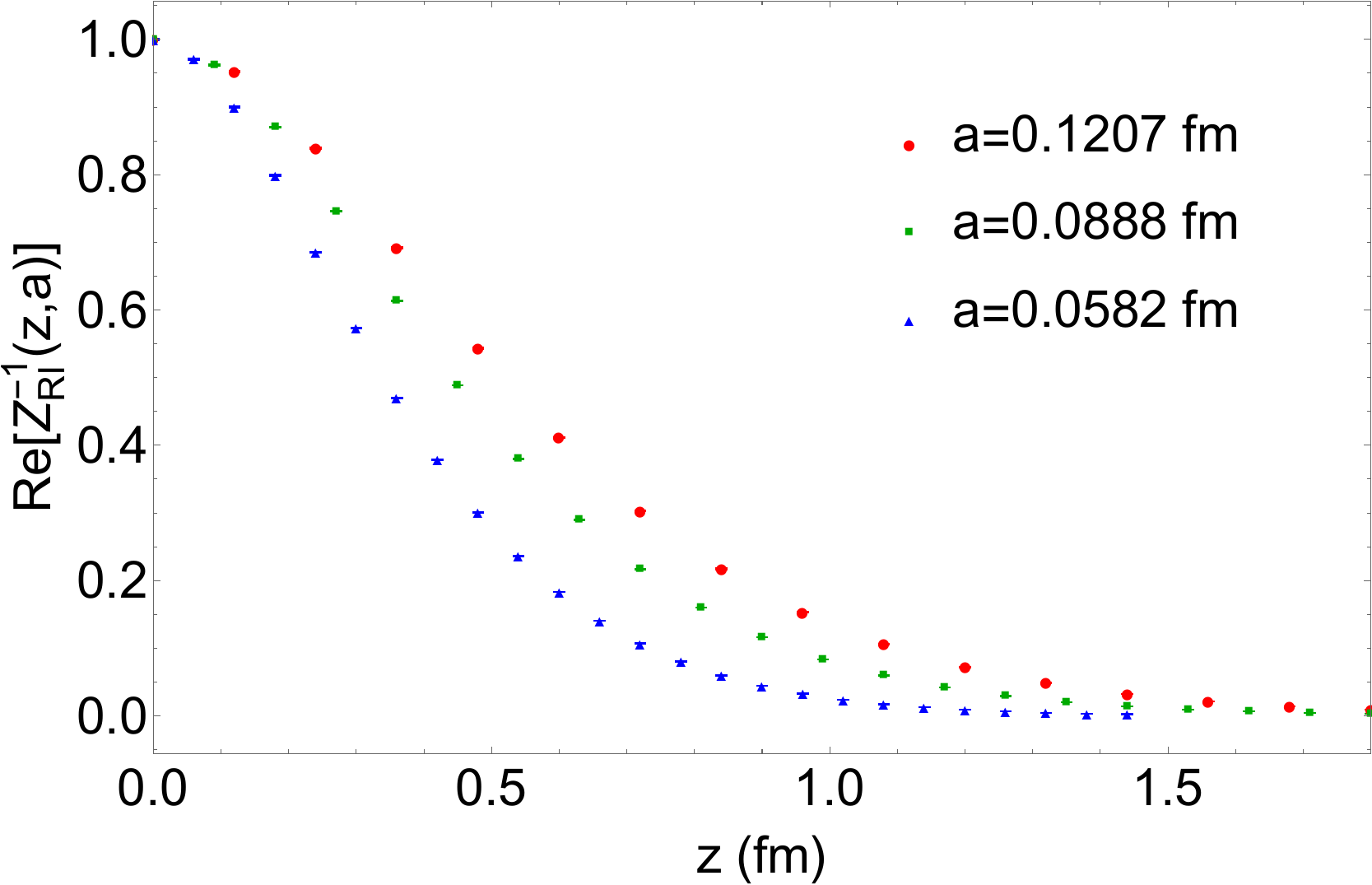}}
  \subfigure{\includegraphics[width=0.3\linewidth]{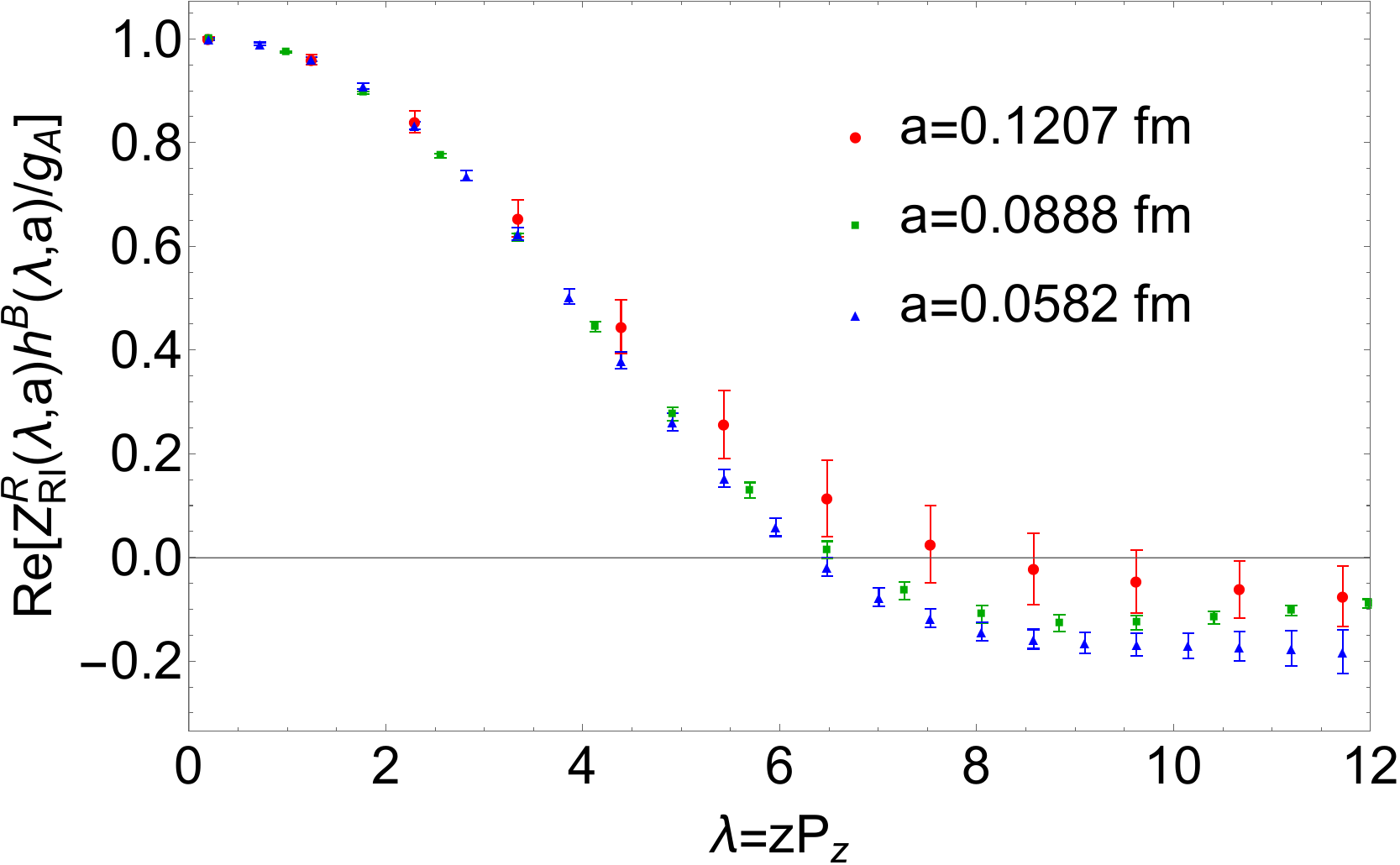}}
  \subfigure{\includegraphics[width=0.3\linewidth]{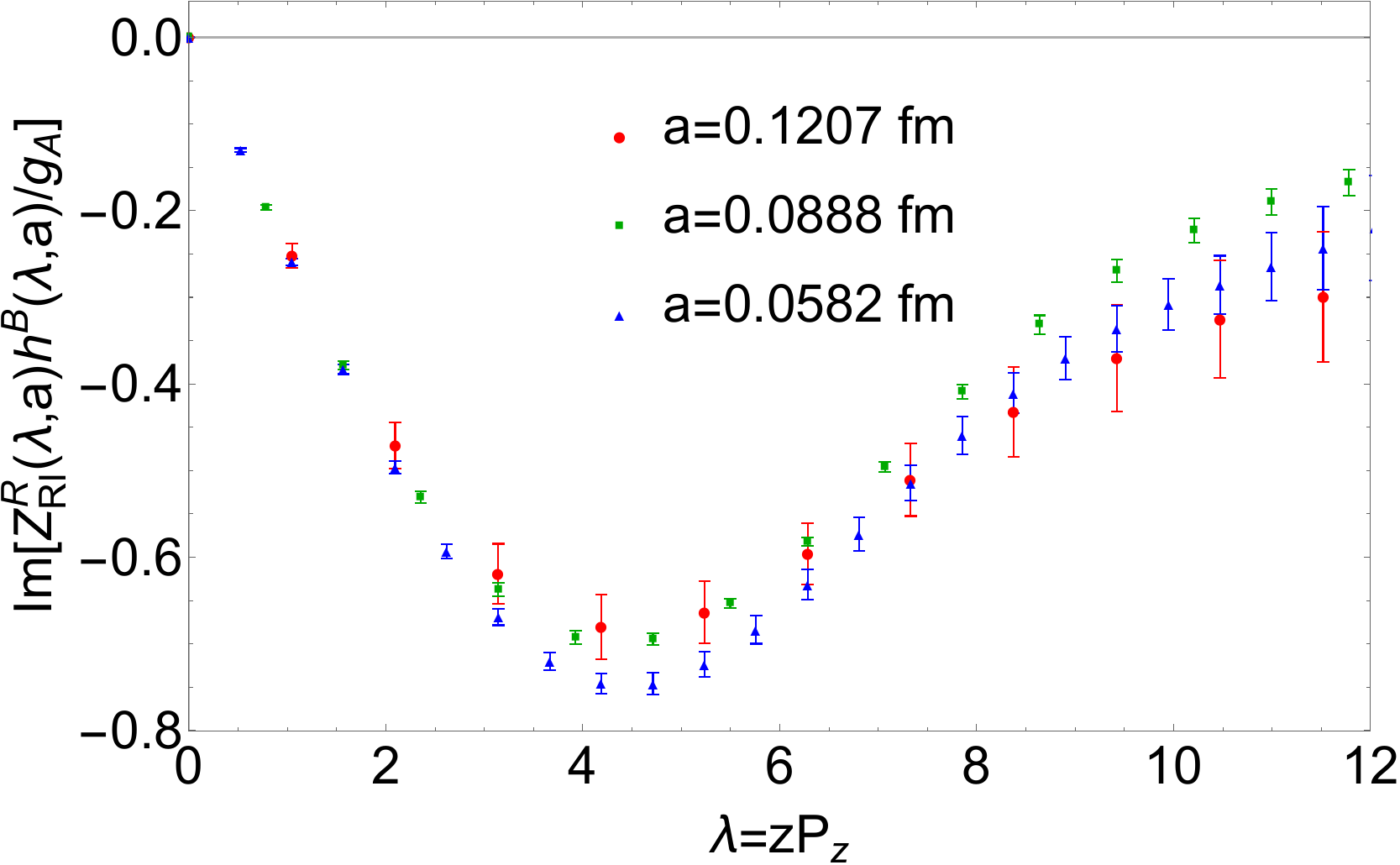}}
  \caption{RI/MOM factors (leftmost figure) at lattice spacings $a\approx\{0.12,0.09,0.06\}$~fm denoted by red circles, green squares and blue triangles, respectively, as a function of Wilson length, $z$; the imaginary parts are compatible with zero. The middle (rightmost) figure shows the real (imaginary) matrix elements renormalized in the pure RI/MOM scheme at the same lattice spacings and the same color scheme as the leftmost plot, as functions of Ioffe time $\lambda=zP_z$.} \label{fig:PureRIMOM}
\end{figure*}

To renormalize the nucleon matrix elements for helicity PDFs, the renormalization factors in the RI/MOM scheme~\cite{Martinelli:1994ty} are calculated in Ref.~\cite{Zhang:2020gaj}.  
The real renormalization factors   as a function of $z$ are shown in 
the leftmost panel of Fig.~\ref{fig:PureRIMOM} while 
the imaginary ones are consistent with zero for $p_z^R=0$ with lattice spacings $a\approx\{0.12,0.09,0.06\}$~fm denoted by red circles, green squares and blue triangles, respectively.
We can see that the RI/MOM factor decays more quickly as $a$ decreases across the whole $z$ range;
this is due to the linear divergence becoming more severe as lattice spacing decreases.
The same behavior was also observed by ETMC~\cite{Alexandrou:2020qtt},
in which the inverse renormalization factors in the RI$^{\prime}$-MOM scheme for the helicity operator were calculated as a function of $z$ at lattice spacings $0.09$, $0.08$ and $0.06$~fm.
This linear divergence can be quantified as the $kz/a\ln(a\LambdaQCD)$ term mentioned in Eqs.~\ref{eq.functionalFormNLO} and \ref{eq.functionalFormNLORGRLRR}
in Sec.~\ref{sec:SelfRenormalization}, which becomes larger as $a$ decreases at fixed $z$.
Later in this work, we will remove this divergence with the HSR procedure. 
In Fig.~\ref{fig:PureRIMOM}, we show the matrix elements renormalized in the pure-RI/MOM scheme with lattice spacings $a\approx\{0.12,0.09,0.06\}$ fm shown as red circles, green squares and blue triangles, respectively, with the real (imaginary) part in the middle (rightmost) panel. 
Our renormalized matrix elements are also normalized to 1 at $z=0$. This is equivalent to dividing the renormalized matrix elements by the axial charge $g_A$ which can be determined both on the lattice and experimentally. 

LaMET calculations use matrix elements boosted to large momentum, and we renormalize them in HSR~\cite{LatticePartonCollaborationLPC:2021xdx}.
The method involves fitting bare matrix elements at $P_z=0$ to a functional form dictated by perturbative QCD and the fitting parameters determined in this process can then be used to divide out the sources of divergence in the large-momentum matrix elements relevant for LaMET.
The case of \N\ can be supplemented with the systematic improvements of RGR and LRR, which we denote by \NLR.
The renormalization factor $\mathcal{Z}(z,a)$ fitted to the functional form can either be an RI/MOM factor $Z_\text{RI}(z,a,\mu_R,p_z^R=0)$ or a hadron matrix element evaluated at zero boost momentum. 
The two functional forms are equally valid, as was demonstrated in Ref.~\cite{Holligan:2024umc}, where the hybrid-RI/MOM and hybrid-ratio schemes were compared. The $m_0$ parameter was extracted in the same way for both schemes by demanding the short-distance behavior ($z\lesssim 0.3$~fm) agree with perturbation theory. It was demonstrated that the renormalized matrix elements are compatible well within error bars between these two schemes. This shows that we can choose either an RI/MOM factor or a zero--boost-momentum matrix element without introducing additional systematic uncertainties.
In our calculation, we choose the RI/MOM factor.
In addition, the matching kernels coincide at \N\ for the hybrid-ratio scheme and the hybrid-RI/MOM scheme for RI/MOM matrix elements evaluated at zero-momentum, as was shown in Ref.~\cite{Chou:2022drv}.
We follow the procedure of Refs.~\cite{LatticePartonCollaborationLPC:2021xdx,LatticeParton:2022xsd,LatticeParton:2022zqc} for \N\ and of Ref.~\cite{Holligan:2023rex} for \NLR.

We describe, first, the self-renormalization procedure at \N.
The logarithm of the renormalization factors is fitted to the functional form \cite{LatticePartonCollaborationLPC:2021xdx}
\begin{multline}\label{eq.functionalFormNLO}
    \left.\ln\left(\frac{1}{\mathcal{Z}(z,a)}\right)\right|_\text{\N} =
    \frac{kz}{a\ln(a\LambdaQCD)} + g_1(z) + f_1(z)a +\\
    \frac{3C_F}{4\pi\beta_0}\ln\left(\frac{\ln(1/a\LambdaQCD)}{\ln(\mu/\LambdaQCD)}\right) +
    \ln\left(1+\frac{d}{\ln(a\LambdaQCD)}\right),
\end{multline}
where $z$ is the Wilson length;
$k$ is the linear divergence arising from the self-energy of the Wilson link.
The linear divergence contains a renormalon ambiguity~\cite{LatticePartonCollaborationLPC:2021xdx} which is accounted for later in the HSR procedure.
$a$ is the lattice spacing;
$\LambdaQCD$ is the cutoff scale for QCD;
$f_1(z)$ is a function describing the discretization effects of the lattice.
The ensembles generated in Ref.~\cite{MILC:2012znn} use clover-improved action for the valence quarks and highly improved staggered quarks (HISQ) for the sea, so the discretization terms are $\mathcal{O}(a)$.
$\mu$ is the final desired renormalization scale;
$d$ is a parameter determined by demanding that the short-distance behavior ($z\lesssim 0.3$~fm as suggested by Ref.~\cite{Ji:2020brr}) of the renormalized matrix element agrees with perturbation theory;
$C_F$ is the quadratic Casimir for the fundamental representation of SU(3) and $\beta_0$ is the first coefficient of the QCD beta-function.
The expression $g_1(z)$ is the residual term that describes the non perturbative physics after the divergences and discretization effects are removed.
We, thus, have an expression for the bare matrix element from which the different sources of divergence can be divided out.

We interpolate the renormalization factors of Wilson length $z\in[0, 1.20]$~fm in uniform steps of $0.06$~fm for all three ensembles.
Next, we determine the linear divergence and QCD scale ($k$ and $\LambdaQCD$) by setting the parameter $d$ to zero initially and fitting to Eq.~\ref{eq.functionalFormNLO} for multiple pairs of $k$ and $\LambdaQCD$. 
For this work, we choose $\LambdaQCD=0.2$~GeV at minimum $\chi^2=1.04$ at $k=0.795$ when $d=0$.

The next step is to remove the logarithmic divergence and renormalon divergence
by setting the short-distance renormalized matrix elements to agree with the Wilson coefficient $C^\text{\N}_0(z,\mu)$.
The helicity Wilson coefficients have been computed to \N~\cite{Yao:2022vtp,Izubuchi:2018srq}: $C_0^\text{\N}(z,\mu) = 1 + \frac{\alpha_s(\mu)C_F}{4\pi}\left(3 \ln\left( \frac{z^2 \mu^2 e^{2\gamma_E}}{4}\right)+7\right)$, where $\alpha_s(\mu)$ is the strong coupling at energy scale $\mu$, and $\gamma_E$ is the Euler-Mascheroni constant.
Since we tabulated the RI/MOM factors in steps of $z=0.06$~fm, we fit to perturbation theory in the interval $z\in[0.06,0.18]$~fm.
We demand that $g_{1,2}(z)-\ln(C^\text{\N}_0(z,\mu))=m_0z+c$ (a linear function) whose $y$-intercept is less than $10^{-3}$ to ensure good matching to perturbation theory.
We tune the $d$ parameter to match $C_0^\text{\N}(z,\mu)$ and we find the value $d=0.497$ gives a $y$-intercept of $\mathcal{O}(10^{-5})$.
We now have a minimum $\chi^2=1.02$ at $k=0.798$, which is a change of less than $0.5\%$ in the value of $k$ compared to $d=0$. We then construct a full renormalization factor 
\begin{multline}
    \frac{1}{Z^\text{\N}_\text{self}(z,a)}=\exp\left[\frac{kz}{a\ln(a\LambdaQCD)}+m^\text{\N}_0z+f_1(z)a+\right.\\
    \left.\frac{3C_F}{4\pi\beta_0}\ln\left(\frac{\ln(1/a\LambdaQCD)}{\ln(\mu/\LambdaQCD)}\right)+\ln\left(1+\frac{d}{\ln(a\LambdaQCD)}\right)\right].
\end{multline}

To perform HSR at \NLR, we first modify the fitting function to~\cite{Holligan:2023rex}
\begin{multline}\label{eq.functionalFormNLORGRLRR}
    \left.\ln\left(\frac{1}{\mathcal{Z}(z,a)}\right)\right|_\text{\NLR} = \Delta\mathcal{I}+\frac{kz}{a\ln(a\LambdaQCD)} + g_2(z) \\
    + f_2(z) a + \frac{3C_F}{4\pi\beta_0}\ln\left(\ln\left(\frac{1}{a\LambdaQCD}\right)\right) +
    \ln\left(1+\frac{d}{\ln(a\LambdaQCD)}\right).
\end{multline}
where the $\mu$ dependence has been removed since it will be resummed with the RGE and we include the conversion constant $\Delta\mathcal{I}$.
We use the same values for $k$, $\LambdaQCD$ and $d$ from the \N~case as was done in Ref.~\cite{Holligan:2023jqh} since the first two are global parameters that depend on the bare matrix elements regardless of whether we use RGR or LRR. This time, the conversion constant is tuned so as to match the short-distance behavior with $C_0^\text{\NLR}(z,\mu)$ Wilson coefficient which is defined in Ref.~\cite{Zhang:2023bxs}. We find $\Delta\mathcal{I}=-0.330$ has a corresponding $y$-intercept of $\mathcal{O}(10^{-3})$. The full self-renormalization factor for \NLR~is
\begin{multline}\label{eq.ZRfactorNLORGRLRR}
    \frac{1}{Z^\text{\NLR}_\text{self}(z,a)} = \exp\left[\Delta\mathcal{I} +
    \frac{kz}{a\ln(a\LambdaQCD)} + m^\text{\NLR}_0z\right. \\
    \left.+f_2(z)a+\frac{3C_F}{4\pi\beta_0}\ln\left(\ln\left(\frac{1}{a\LambdaQCD}\right)\right) +
    \ln\left(1+\frac{d}{\ln(a\LambdaQCD)}\right)\right].
\end{multline}

We perform the continuum extrapolation on the quantity $Z^\text{X}_\text{self}(z,a)h^B(z,a)$, where X is \N~or \NLR~and finally convert to the hybrid scheme with self-renormalization with the factor
\begin{equation}
    Z_\text{HSR}^\text{X}(z,a) = \frac{Z^\text{X}_\text{self}(z,a)}{C^\text{X}_0(z,\mu)}\theta(z_s-z) + \frac{Z^\text{X}_\text{self}(z,a)}{C^\text{X}_0(z_s,\mu)}\theta(z-z_s);
\end{equation}
where $z_s\approx 0.3$~fm is the maximum distance at which perturbation theory is valid. Thus, the full hybrid-renormalized matrix element is $h^{R,\text{X}}_\text{HSR}(z,a) = Z^\text{X}_\text{HSR}(z,a)h^B(z,a)$.

\begin{figure*}[ht]
  \centering
  \subfigure{\includegraphics[width=0.3\linewidth]{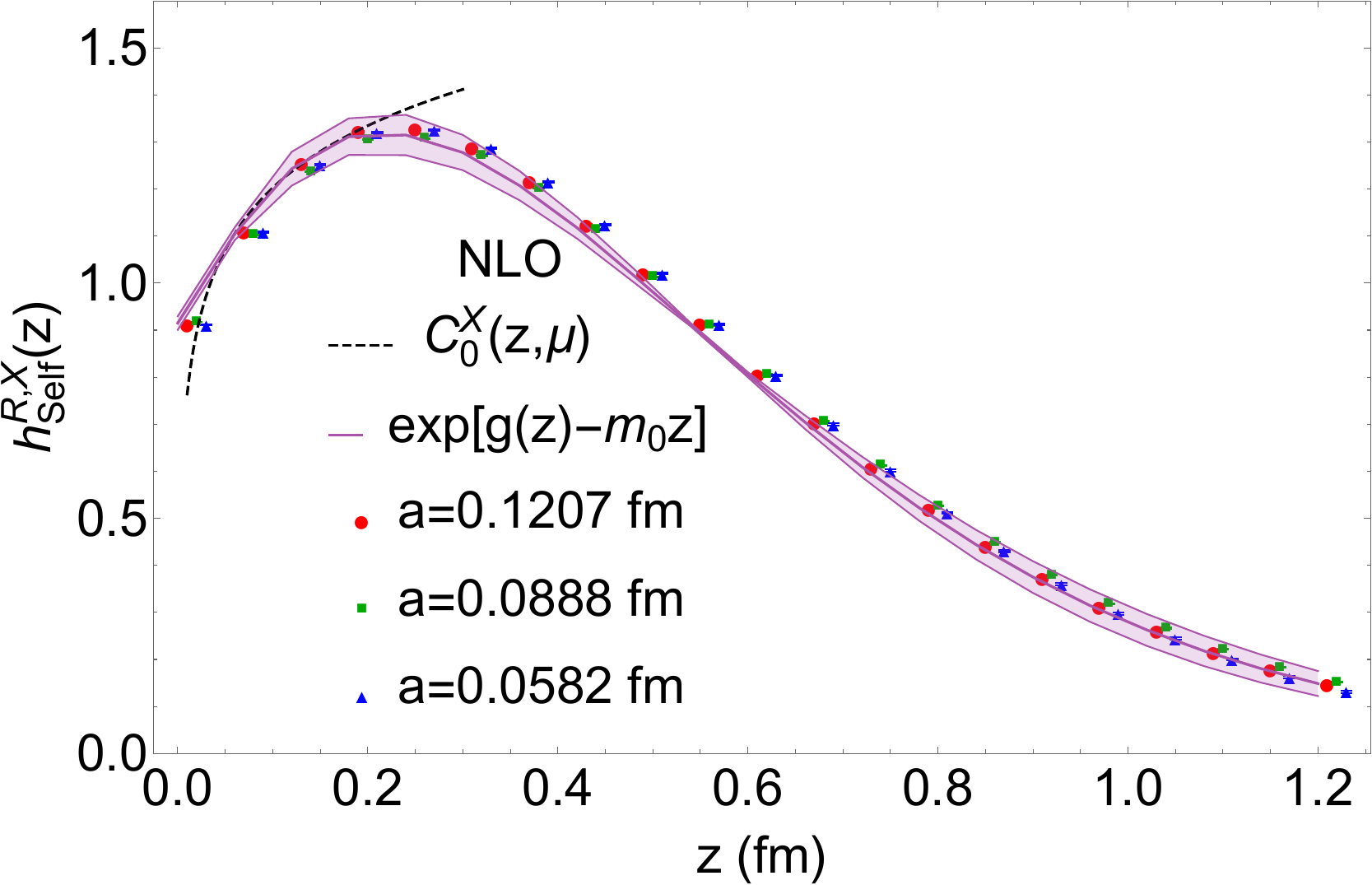}\quad
  \subfigure{\includegraphics[width=0.3\linewidth]{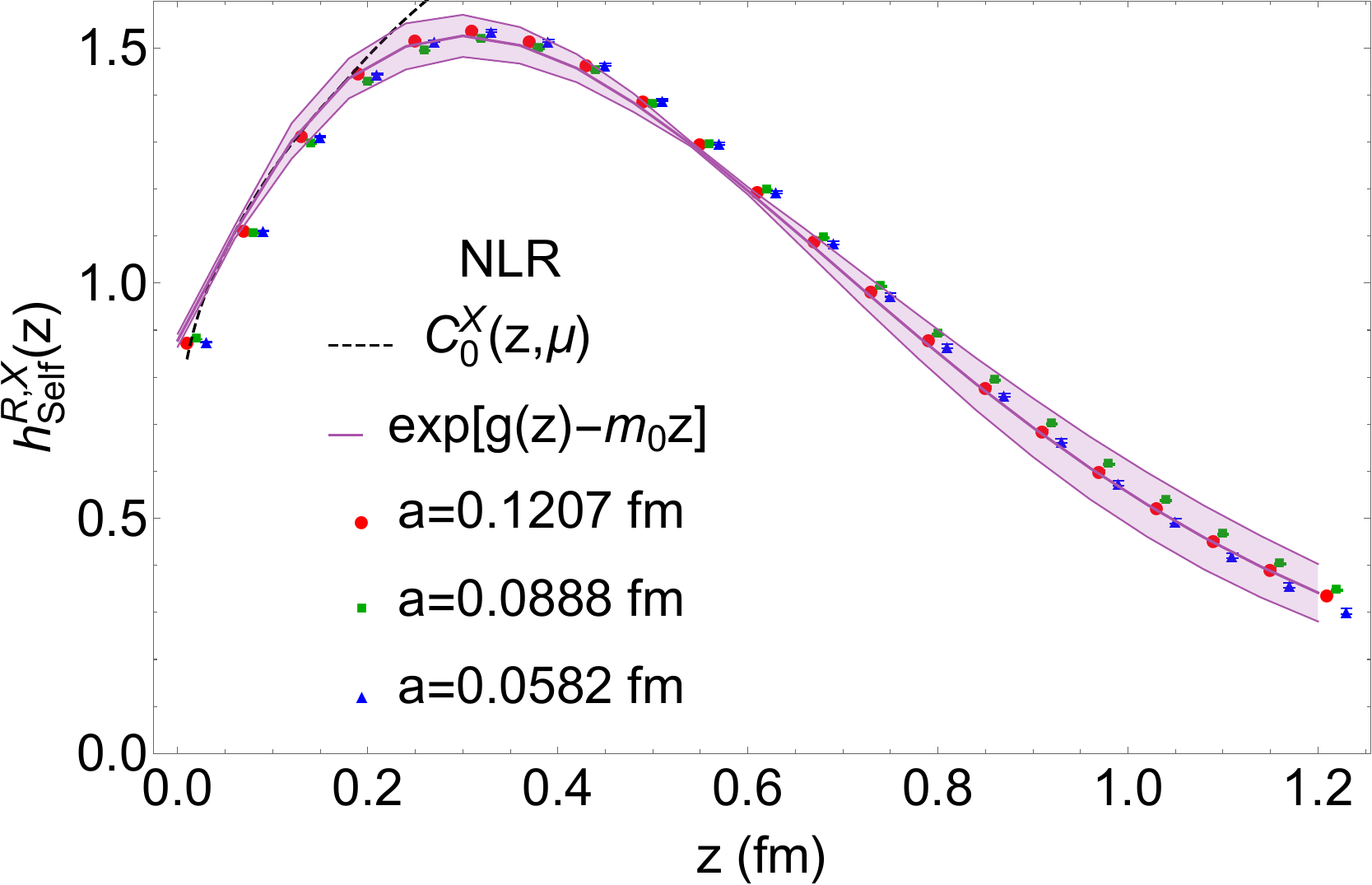}}}
    \caption{$h^R_\text{self}(z,a)$ at lattice spacings $a\approx\{0.12,0.09,0.06\}$~fm are shown as red circles, green squares and blue triangles, respectively;
    the left (right) plot shows \N\ (\NLR).
    All but the largest $a$ have been offset slightly to the right from their true $z$ value to allow for readability.
    We also plot the Wilson coefficient $C^\text{X}_0(z,\mu)$ as a dotted black line, which agrees with the renormalized data for short distances.
    The purple band shows the term $\exp(g(z)-m_0z)$.}
    \label{fig:SelfRenormalization}
\end{figure*}

The matrix elements renormalized in the self-renormalization scheme, $h^{R,\text{X}}_\text{self}(z)$, for both X~being~\N\ and \NLR\ are shown in Fig.~\ref{fig:SelfRenormalization} to demonstrate the effectiveness of self-renormalization.
We show the renormalization factors at $a\approx\{0.12,0.09,0.06\}$~fm in red circles, green squares and blue triangles, respectively, and the corresponding Wilson coefficient as a black-dotted line.
Note that the dependence on lattice spacing $a$ is almost completely removed by the self-renormalization process.
The results at the smallest and largest lattice spacings differ by no more than 6\% for $z\in[0,1.0]$~fm.
In addition, the matrix elements agree with the corresponding Wilson coefficient for short distances, plotted as a black dashed line.
The systematic errors are estimated using the method of ``scale variation'' as in Refs.~\cite{Zhang:2023bxs,Holligan:2023jqh}.
When we use the RGR for the matrix elements, we set the initial scale $\mu=\mathtt{z}^{-1}$ so as to eliminate the logarithms and then evolve to the final desired energy scale $\mu=2.0$~GeV.
Scale variation involves setting the initial scale to $c'\times\mathtt{z}^{-1}$ for $c'=0.75$ and $c'=1.5$, as was used in Ref.~\cite{Zhang:2023bxs};
the central value corresponds to $c'=1.0$.
This corresponds to a variation of approximately $15\%$ on either side of $\alpha_s(\mu=2.0\text{ GeV})$ in the strong coupling.

The first thing we notice is that there is a significant decrease in systematic errors when going from \N\ to \NLR\ for all lattice spacings in the real and imaginary parts.
The relative systematic errors for the real part decrease by up to a factor of eight for $a=0.1207$~fm and $a=0.0582$~fm, and up to a factor of seventeen at $a=0.0888$~fm.
The effect on the imaginary part is even greater with a decrease of as much as a factor of twenty for all lattice spacings.

\begin{figure*}[ht]
  \centering
  \subfigure{\includegraphics[width=0.3\linewidth]{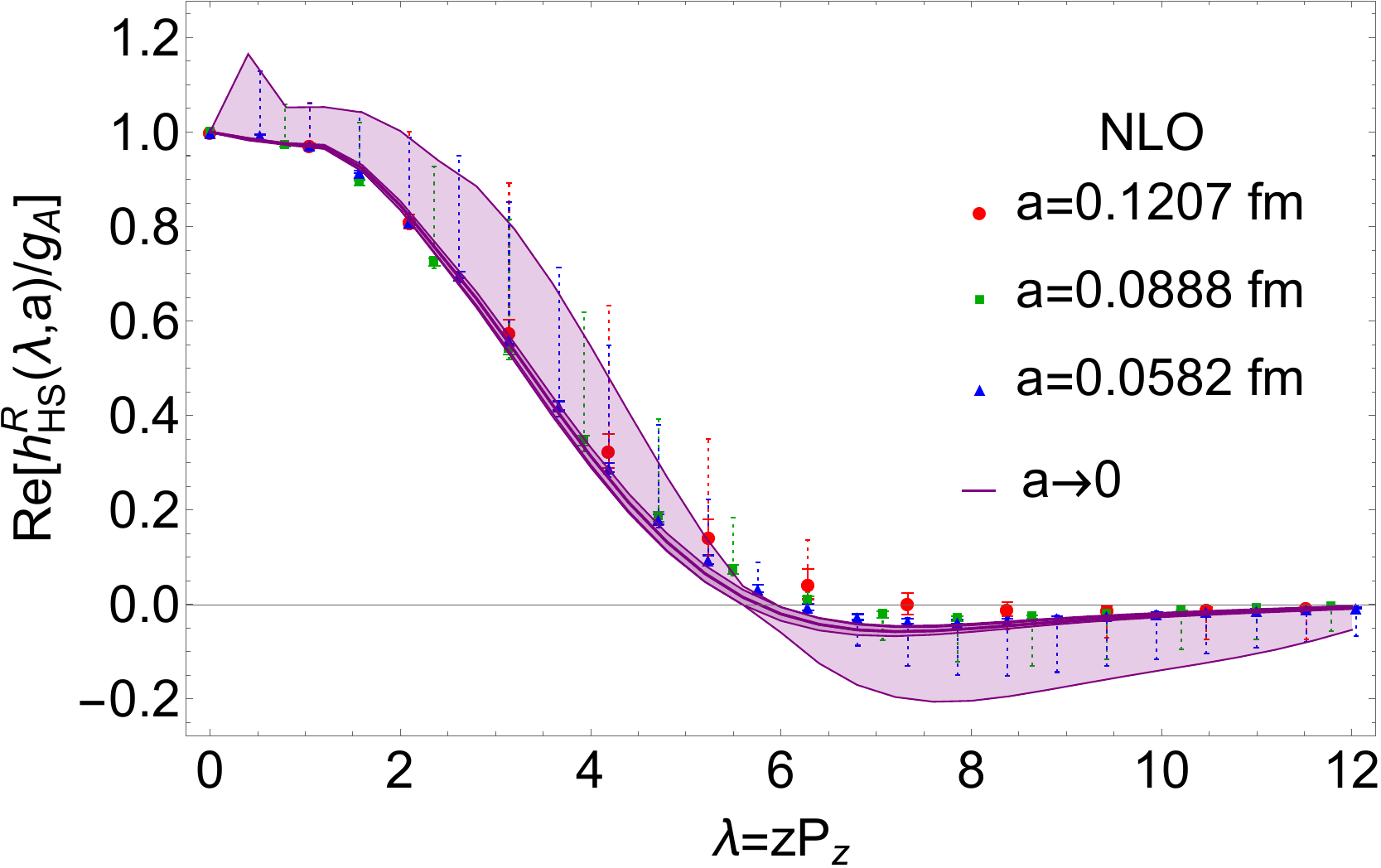}
  \subfigure{\includegraphics[width=0.3\linewidth]{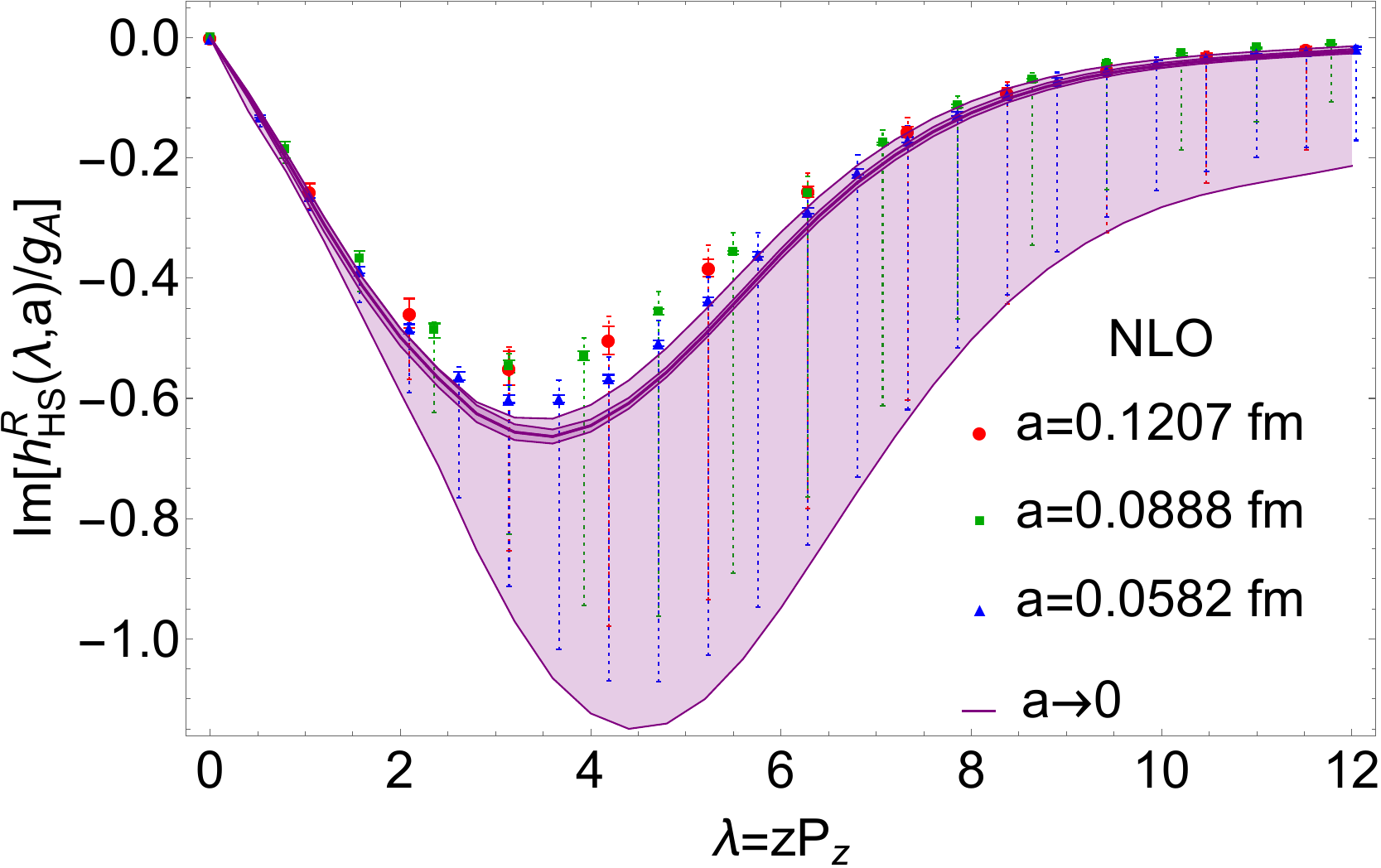}}}
  \subfigure{\includegraphics[width=0.3\linewidth]{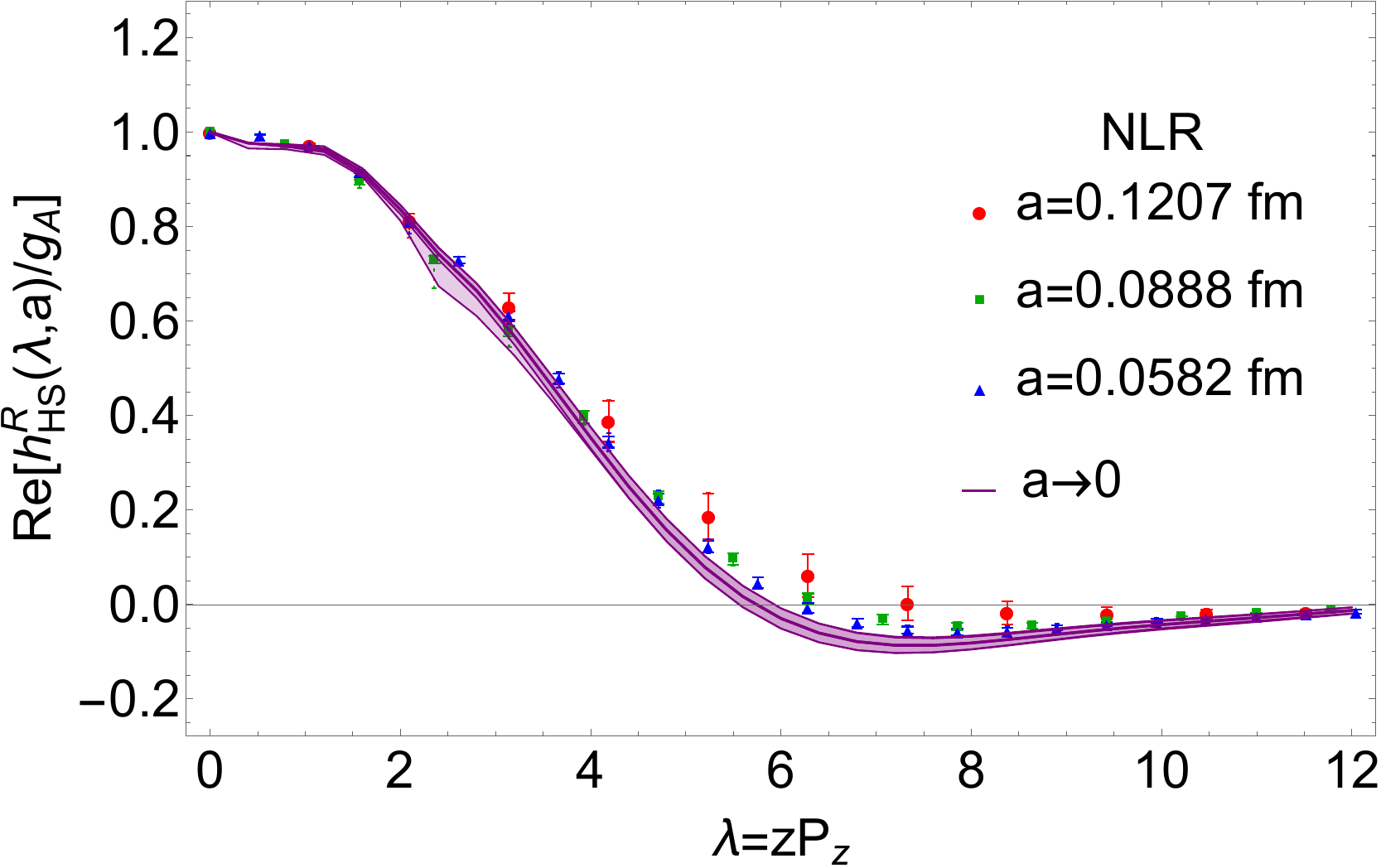}
  \subfigure{\includegraphics[width=0.3\linewidth]{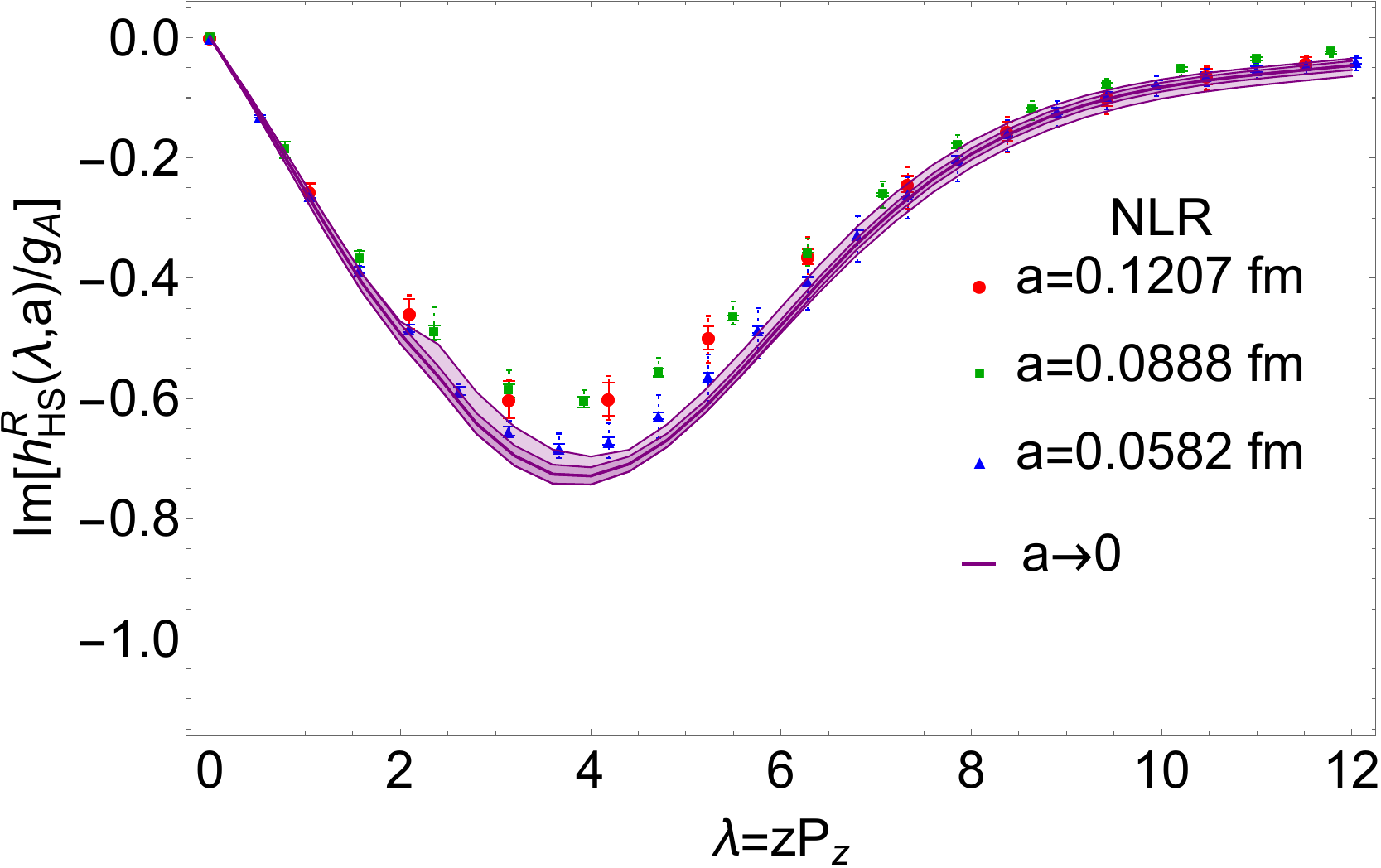}}}
  \caption{Real (left column) and imaginary (right column) matrix elements renormalized in the HSR at lattice spacings $a=\{0.1207,0.0888,0.0582\}$~fm and the continuum limit plotted in red, green, blue and purple, respectively.
  Data are shown at both \N~(top row) and \NLR~(bottom row).
  The continuum extrapolation is computed by performing a weighted fit to a function linear in $a$ of the discrete data at fixed $z$-value.
  The solid error bars are statistical and the dotted error bars are combined statistical and systematic errors the latter of which are derived from scale variation as described earlier in this section.}\label{fig:hR_SR_sys_variable_a}
\end{figure*}

The continuum extrapolation is performed on the quantity $h^{R,\text{X}}_\text{self}(\lambda,a)$ by fitting to a linear function $h^{R,\text{X}}_\text{self}(\lambda,a)=c(\lambda)\times a + h^{R, \text{X}}_\text{self}(\lambda,a=0)$ where $\lambda=zP_z$ is Ioffe time.
In the continuum case, the relative systematic errors for the real part decrease by as much as a factor of six when NLO is supplemented with both RGR and LRR;
the same quantities for the imaginary part decrease by as much as a factor of eight.
This is the same effect observed in Ref.~\cite{Holligan:2023jqh}, in which the systematic errors are drastically reduced with the RGR and LRR improvements, as well as the fact that the large logarithms and renormalon divergence are significant sources of systematic errors.
In Fig.~\ref{fig:hR_SR_sys_variable_a} we show the renormalized matrix elements at fixed X and variable $a$.
The top (bottom) row shows the hybrid-renormalized matrix elements at \N\ (\NLR) while the left (right) column shows the real (imaginary) part with those from lattice spacings $a=\{0.1207,0.0888,0.0582\}$~fm shown in red circles, green squares and blue points
and continuum-extrapolated ones shown as a purple band, respectively.
We show both statistical errors (solid inner lines) and statistical and systematic errors combined with quadrature (dashed outer lines).
We also see that the relative systematic errors vary by as much as a factor of two between the largest and smallest lattice spacings for both \NLR\ and \N\ in the real and imaginary parts.
The scale of the systematics at \N\ and \NLR\ affect the systematics of the corresponding continuum extrapolation.
The smaller systematic errors in the $a\neq 0$ data at \NLR\ compared to \N\ results in smaller systematics in the continuum extrapolation.
Comparing both \N\ and \NLR\ renormalized matrix elements from $a=0.0888$~fm and $0.0582$~fm (the results with the smallest statistical errors), we see that the real parts differ by at most one sigma across the full range of $\lambda$.
The imaginary parts show some tension in the region $\lambda\in[4,6]$ but elsewhere are also compatible within two sigma.
By contrast, the matrix elements renormalized in the pure-RI/MOM scheme shown in Fig.~\ref{fig:PureRIMOM} show compatibility for both real and imaginary parts within two sigma but have much larger error bars than the matrix elements renormalized with the HSR.
In the continuum limit, the real part of the central values in the region $\lambda\in[0,4]$ differ by up to 10\% with those at $a=0.0582$~fm and up to 20\% with those at $a=0.1207$~fm.
The imaginary part of the central values in the same $\lambda$ range differ by no more than 5\% with those at $a=0.0582$~fm and 7\% with those at $a=0.1207$~fm.
Both of these demonstrate good convergence in the continuum limit.

\section{Helicity Parton Distribution Function}

To obtain the nucleon isovector helicity PDF, we first need to Fourier transform the HSR matrix elements $h^{R,\text{X}}_\text{HSR}(z,a)$ to momentum space to obtain the quasi-PDF $\Delta\tilde{q}^\text{X}(x,P_z)=\int^{\infty}_{-\infty}\frac{P_z\diff{}{z}}{2\pi}e^{ixzP_z}h^{R,\text{X}}_\text{HSR}(z,a)$.
However, simply truncating the integral at the limit of the Wilson length extent on the lattice will cause unphysical oscillations in the quasi-PDF~\cite{Ji:2020brr,Gao:2021dbh,Chen:2018xof}.
To prevent this, we extrapolate the renormalized matrix elements to infinite distance.
The large-distance behavior of the renormalized matrix element determines the small-$x$ behavior of the lightcone PDF; the extrapolation model for infinite distance is derived from a model assumption $x^{k_1}$, commonly used in the global-fit community, for naive small-$x$ region of PDFs.
We adopt the large-$z$ extrapolation model previously used in Refs.~\cite{Ji:2020brr,Holligan:2023jqh,Gao:2021dbh,Gao:2022uhg}:
$   h^{R, \text{X}}_\text{HSR}(z,a)/g_A \to \frac{Ae^{-mz}}{|zP_z|^n}\text{~~~as $z\to\infty$}$,
where $A$, $m$ and $n$ are fitting parameters.
The range of $z$ values used must be large enough to realistically capture the long-distance behavior but not so large that the matrix elements are too noisy.
In all cases, the $\chi^2/\rm{dof}$ value is less than 1 which demonstrates that the large distance behavior is well captured by the fitting model.
We can then use combined renormalized matrix elements directly from the lattice
and those large-$z$ extrapolation to Fourier transform to obtain quasi-PDFs.

The final stage in the calculation is the perturbative matching which is used to align the UV behavior of the quasi-PDF with the lightcone.
The lightcone PDF is related to the quasi-PDF via $\Delta q^\text{X}(x,\mu) = \int_{-\infty}^{\infty}\frac{\diff{}{y}}{|y|}\mathcal{K}^{-1, \rm X}(x,y,\mu,P_z,z_s)\Delta\tilde{q}^\text{X}(y,P_z)+\mathcal{O}\left(\frac{\LambdaQCD^2}{P_z^2x^2(1-x)}\right)$
where $\mathcal{K}^{-1,\rm X}$ is the matching kernel for case X, which has been derived up to one-loop for helicity PDFs renormalized in the hybrid scheme~\cite{Chou:2022drv,Yao:2022vtp} with finite-$P_z$ corrections~\cite{Gao:2021dbh,Braun:2018brg}.
To make the matching kernel at \NLR, we first add the LRR modification term $\Delta\mathcal{K}^\text{LRR}$ defined in Refs.~\cite{Holligan:2023rex,Zhang:2023bxs}
to the matching kernel to make $\mathcal{K}^\text{\NL}=\mathcal{K}^\text{\N}+\Delta\mathcal{K}^\text{LRR}$.
The process of RGR in the matching follows the same philosophy as in the case of renormalization.
The matching is performed with the kernel $\mathcal{K}^\text{\NL}$ at energy scale $\mu=2xP_z$ (such that the scale dependence vanishes from matching formula).
We then evolve the matched PDF to the final desired energy scale (in our case, $\mu=2.0$~GeV), this time using the Dokshitzer-Gribov-Lipatov-Altarelli-Parisi (DGLAP) equation:
$  \frac{\diff{}{\Delta q^\text{X}(x,\mu)}}{\diff{}{\ln(\mu^2)}} =
  \int^1_x \frac{\diff{}{z}}{|z|} \mathcal{P}(z) \Delta q^\text{X}\left(\frac{x}{z},\mu\right)$,
where $\mathcal{P}(z)$ is the DGLAP kernel, which has been calculated up to three
loops~\cite{Moch:2004pa}.
It should be noted that the DGLAP evolution formula begins to break down at $|x|\lesssim 0.2$ where $\alpha_s(\mu=2xP_z)$ becomes nonperturbative.
We use the same algorithm for RGR matching as was detailed in Appendix~D of Ref.~\cite{Su:2022fiu}.

\begin{figure*}[ht]
  \centering
    \subfigure{\includegraphics[width=0.4\linewidth]{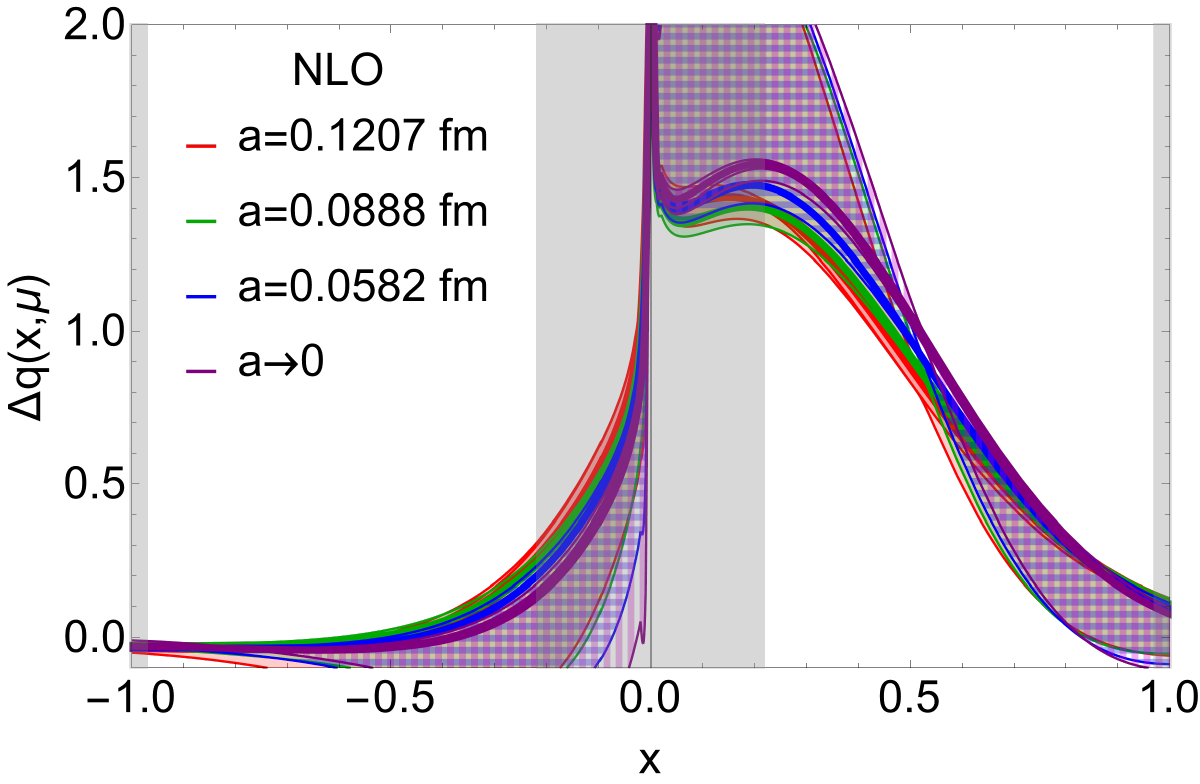}\quad
    \subfigure{\includegraphics[width=0.4\linewidth]{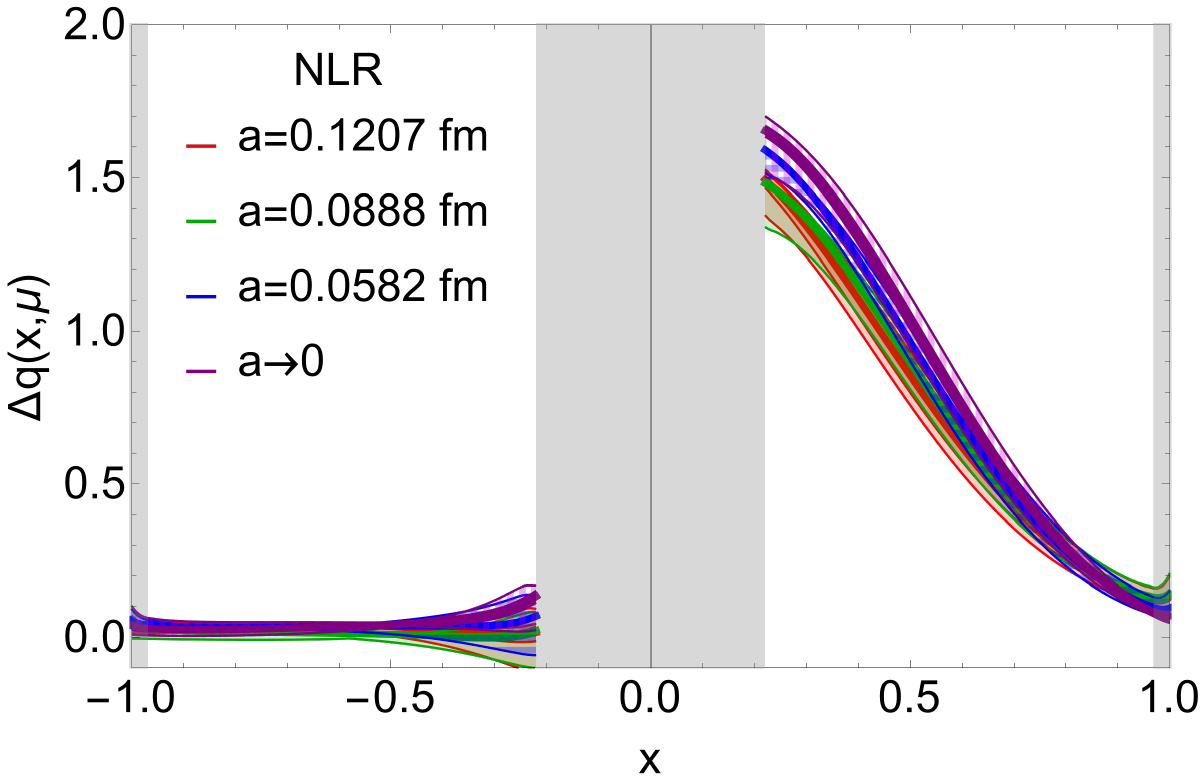}}}
  \caption{Lightcone PDFs at variable lattice spacing with pion mass $M_{\pi}=310$~MeV renormalized in the HSR scheme at \N~(left) and \NLR~(right).
  PDFs from lattice spacings $a=\{0.1207,0.0888,0.0582\}$~fm are plotted in solid-red, solid-green and horizontally hatched blue, respectively.
  The continuum extrapolation is plotted in vertically hatched purple.
  The inner error bars are statistical, and the outer error bars are combined statistical and systematic errors.}\label{fig:PDF_SR_fixed_scheme}
\end{figure*}

We compare the lattice-spacing dependence of  \N\ (left) and  \NLR\ (right) helicity PDFs in Fig.~\ref{fig:PDF_SR_fixed_scheme} with $a=\{0.1207,0.0888,0.0582,0.0\}$~fm bands shown in solid red, solid green, horizontally-hatched blue and vertically-hatched purple, respectively.
We can see that at the \N\ antiquark region has the greatest deviation from zero at $a=0.1207$~fm, the largest lattice spacing, but tends towards zero as lattice spacing decreases, including in the continuum limit.
All \N\ antiquark distributions are compatible with zero with full error bars for $x=[-1,-0.15]$.
By contrast, in the \NLR\ case, the antiquark region does not show dependence on lattice spacing and shows a small positive value.
In the quark region, the relative systematic errors of the \N\ helicity PDF are not correlated with lattice spacing:
the largest systematic error occurs at $a=0.0888$~fm in the interval $x=[0.3,0.5]$ and at $a=0.0582$~fm in the interval $x=[0.5,0.85]$.
For $x=[0.15,0.4]$ the smallest systematic error occurs at the largest lattice spacing.
However, with \NLR\ PDF, both the upper and lower systematic errors decrease with lattice spacing in the range $x=[0.3,0.7]$.
The correlation disappears at small $x$ and large $x$ where the RGR matching and LaMET calculations begin to break down, and the results become unreliable.
This suggests that
the discretization effects are the dominant source of systematics in the \NLR\ case.
We also compare the PDFs at $a=0.1207$~fm and $a=0.0582$~fm (the largest and smallest lattice spacings):
in the interval $x=[0.2,0.8]$, the central values differ by no more than 12\% for both \N\ and \NLR.
The lattice-spacing dependence of the helicity PDF by ETMC~\cite{Alexandrou:2020qtt} with RI$^{\prime}$-MOM and RI-xMOM renormalization schemes shows a greater difference across a narrower range of lattice spacings: $a\in\{0.06,0.08,0.09\}$~fm.
Our own relatively small variation between our largest and smallest lattice spacings compared to the above shows that more of the lattice-spacing dependence is removed by the HSR procedure compared to RI$^{\prime}$-MOM and RI-xMOM.

\begin{figure*}[ht]
  \centering
    \subfigure{\includegraphics[width=0.4\linewidth]{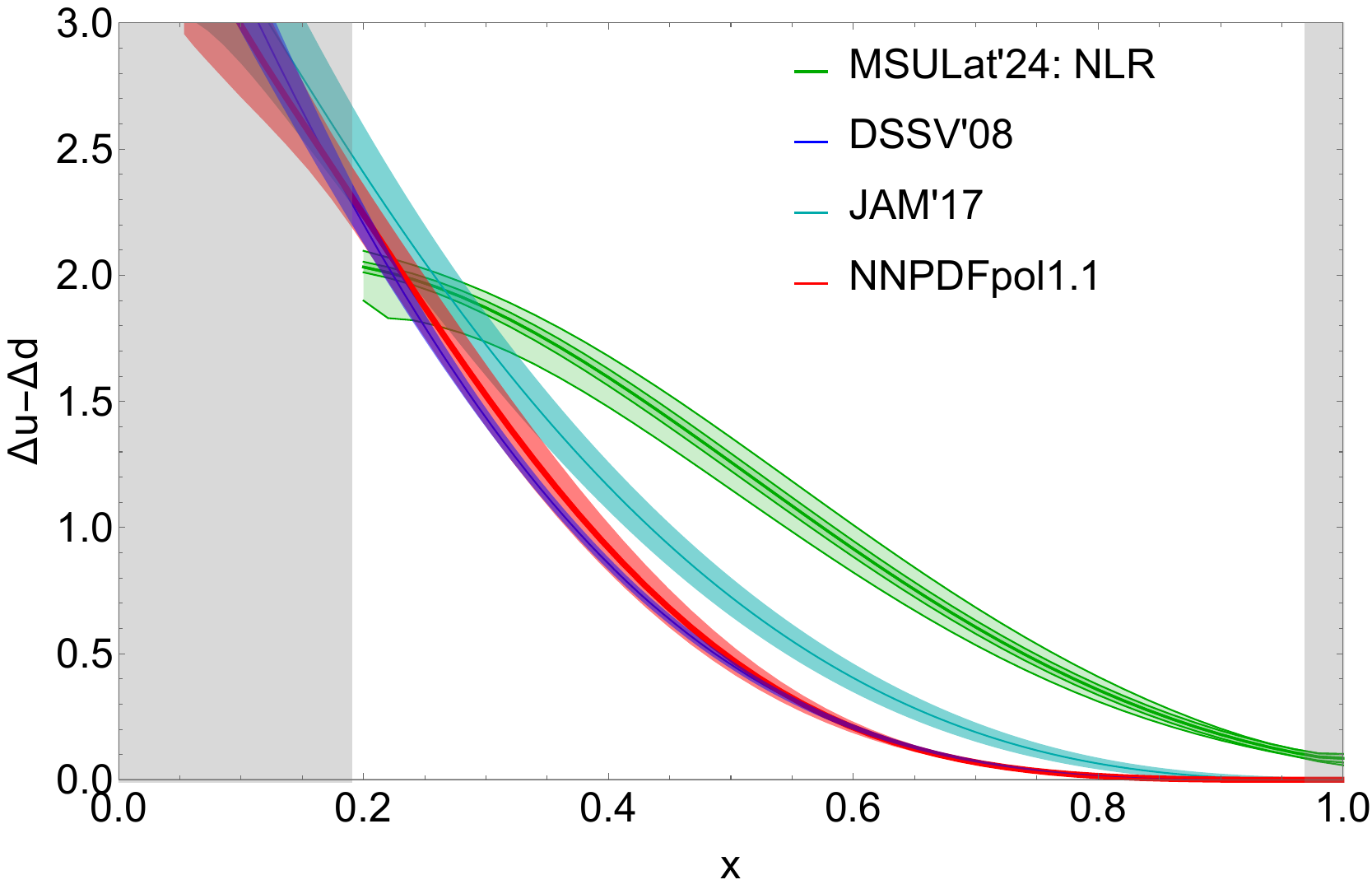}\quad
    \subfigure{\includegraphics[width=0.4\linewidth]{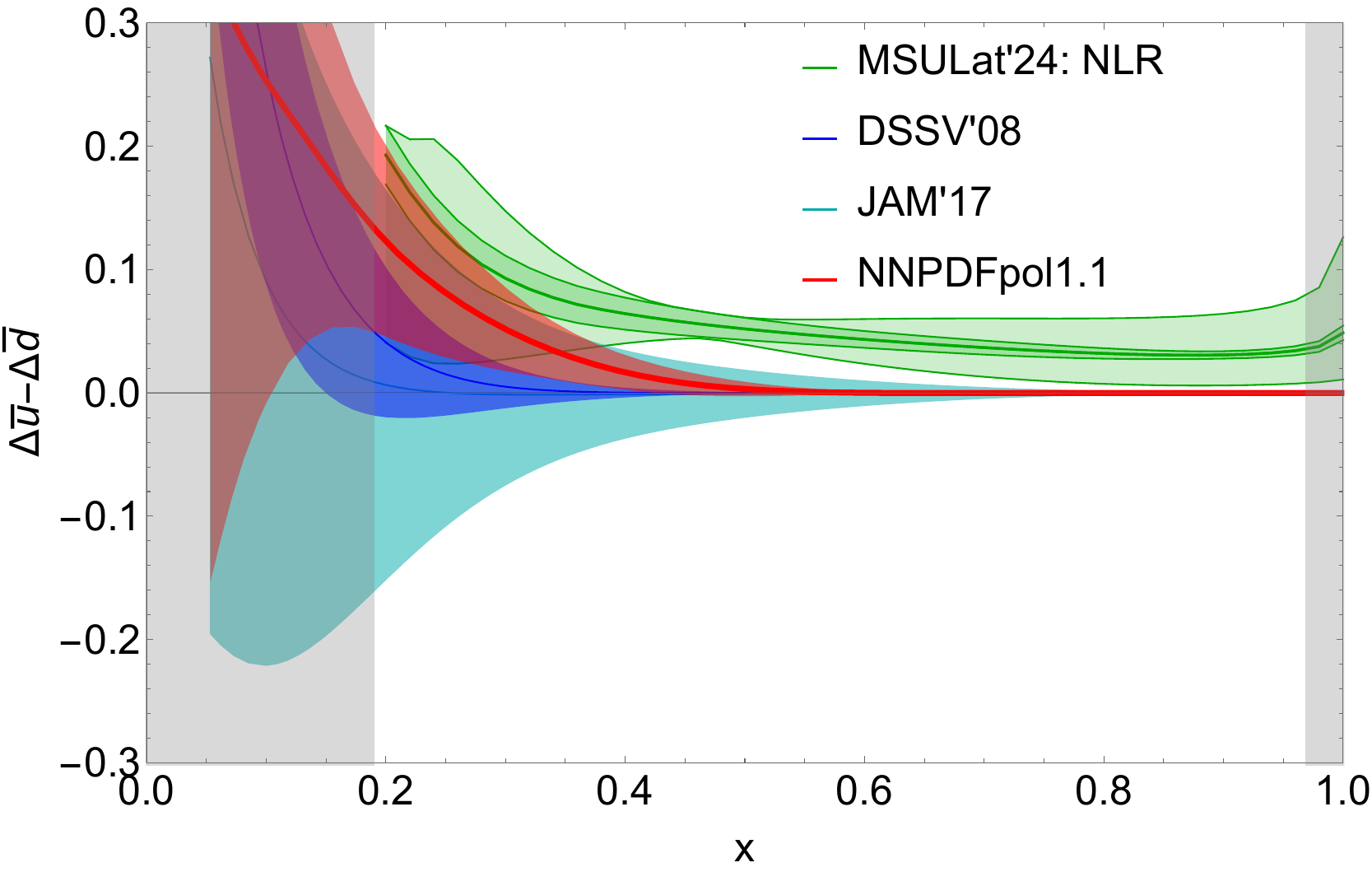}}}
  \caption{Comparison of our $x$-dependent PDFs (``\texttt{MSULat'24}") at \NLR\ (green) with global fits computed in Refs.~\cite{deFlorian:2009vb} (``\texttt{DSSV'08}", blue), \cite{Ethier:2017zbq} (``\texttt{JAM'17}", cyan) and \cite{Nocera:2014gqa} (``\texttt{NNPDFpol1.1}", red).
  The left (right) plot shows the quark (antiquark) region. Our results use the axial charge $g_A=1.218(25)(30)$ from Ref.~\cite{Gupta:2018qil}.}\label{fig:Global_Fit_Comp}
\end{figure*}

In Fig.~\ref{fig:Global_Fit_Comp} we compare our  continuum-limit \NLR\ $x$-dependent PDFs (green)  with the global fits of the NNPDFpol1.1~\cite{Nocera:2014gqa} (red), JAM~\cite{Ethier:2017zbq} (cyan) and DSSV~\cite{deFlorian:2009vb} (blue) collaborations with the quark-region (antiquark-region) shown in the left (right) panel. 
Up until now, our results have been divided by the axial charge $g_A$. The global fits do not use normalization so we multiply our helicity PDFs by the value $g_A=1.218(25)(30)$ computed in Ref.~\cite{Gupta:2018qil}. We use this value since the calculation was performed on the same lattices as our LaMET calculation.
In the quark region, we note that there is tension between the global fits of $\texttt{JAM'17}$ and each of $\texttt{DSSV'08}$ and $\texttt{NNPDFpol1.1}$ from mid to large $x$, which suggests that there are other sources of systematic errors that may have been ignored or underestimated, likely due to the difference in the experimental cuts, theory inputs, parametrization, and so on.
For example, JAM excludes SIDIS data;
DSSV and NNPDFpol1.1 rely on assumptions such as SU(3) symmetry to constrain the analysis and add a very small symmetry-breaking term.
These assumptions are needed due to the difficulties in constraining data from polarized experiments.
Future experiments with neutral- and charged-current DIS (such as at the EIC) will provide useful measurements to constrain our understanding of the antiquark helicity distribution.
Our lattice PDF result also has significant tension with each of the global fits at large $x$ and do not become compatible with zero as $x\to 1$.
This is in contrast to the global fits, which use a $(1-x)^b$ term in the parametrization form, enforcing that the PDF goes to zero as $x\to 1$.
Similar behavior was observed in the past LaMET lattice calculations of the nucleon helicity PDF~\cite{Chen:2016utp,Alexandrou:2020qtt,Alexandrou:2016jqi} at heavy quark mass where no parametrization form is used.
Turning to the antiquark region, we reflect our $x$-dependent PDFs in the vertical to allow a direct comparison with global fits.
Our \NLR\ result for antiquark helicity favors more polarized up than down flavor (whereas \N\ does not).
This agrees with the results of the STAR~\cite{STAR:2014afm} and PHENIX~\cite{PHENIX:2015ade} collaborations which measured $\Delta \overline{u}(x)>\Delta \overline{d}(x)$ and the global fits who use these experimental data as inputs.
The total antiquark flavor asymmetry from this work is $\int_{0.2}^{1}\diff{}{x}(\Delta\overline{u}(x)-\Delta\overline{d}(x))=0.037^{+0.019}_{-0.023}$.

\section{Conclusion}

In this paper we perform the first LaMET calculation of the isovector nucleon helicity PDF using the hybrid scheme with self renormalization and the first lattice-QCD helicity PDF results with the RGR and LRR improvements in both the renormalization and the lightcone matching process.
We use lattice spacings $a=\{0.1207,0.0888,0.0582\}$~fm and perform a continuum extrapolation with a pion mass of $M_{\pi}\approx 315$~MeV.
We demonstrate the use of the hybrid-self renormalization scheme in its application to the isovector nucleon helicity PDF.
We show the fine-tuning procedure for $d$ and $\Delta\mathcal{I}$ to match the renormalized matrix elements to perturbation theory at $z\lesssim 0.2$~fm for \N\ and \NLR, respectively.

Our matrix elements renormalized in the hybrid-self scheme show a reduced dependence on lattice spacing compared to pure-RI/MOM allowing for a more accurate extrapolation to the continuum.
Both renormalized matrix elements and quasi-PDFs have their systematic errors due to scale variation reduced by the LRR and RGR procedures by as much as a factor of 10, but the central values by only about 10\%.
We compared the central values of the \N\ and \NLR\ $x$-dependent helicity PDFs and found that that the application of RGR and LRR has a minor impact on the PDF in the quark region but a significant one in the antiquark region, where it changes sign.
We then use the \NLR\ PDF to compare with global fits, since it is more reliable, due to its accounting for large logarithms and the renormalon divergence.
Our results in quark region do show some tension with the global fits of DSSV'08 \cite{deFlorian:2009vb}, JAM'17 \cite{Ethier:2017zbq} and NNPDFpol1.1 \cite{Nocera:2014gqa}, but so are the JAM'17 and  NNPDFpol1.1, for example.
Our antiquark \NLR\ PDF shows agreement with STAR~\cite{STAR:2014afm} and PHENIX~\cite{PHENIX:2015ade} results in that $\overline{u}(x)>\overline{d}(x)$ and contribute to asymmetry of $\int_{0.2}^{1}\diff{}{x}(\Delta\overline{u}(x)-\Delta\overline{d}(x))=0.037^{+0.019}_{-0.023}$.
Our calculation can be supplemented with future improvements in lattice-QCD calculation systematics for the LaMET method, such as a higher boost momentum and a physical pion mass.

{\bf Acknowledgements: }
We thank the MILC Collaboration for sharing the lattices used to perform this study.
The LQCD calculations were performed using the Chroma software suite~\cite{Edwards:2004sx}.
This research used resources of the National Energy Research Scientific Computing Center, a DOE Office of Science User Facility supported by the Office of Science of the U.S. Department of Energy under Contract No. DE-AC02-05CH11231 through ERCAP;
facilities of the USQCD Collaboration, which are funded by the Office of Science of the U.S. Department of Energy,
and supported in part by Michigan State University through computational resources provided by the Institute for Cyber-Enabled Research (iCER).
The work of JH and HL are partially supported
by the US National Science Foundation under grant PHY 1653405 ``CAREER: Constraining Parton Distribution Functions for New-Physics Searches'', grant PHY~2209424,
by the U.S.~Department of Energy under contract DE-SC0024582,
and by the Research Corporation for Science Advancement through the Cottrell Scholar Award.

\bibliographystyle{ieeetr}

\bibliography{refs}

\begin{thebibliography}{10}

\bibitem{HERMES:2004zsh}
A.~Airapetian {\em et~al.}, ``{Quark helicity distributions in the nucleon for up, down, and strange quarks from semi-inclusive deep-inelastic scattering},'' {\em Phys. Rev. D}, vol.~71, p.~012003, 2005.

\bibitem{COMPASS:2009kiy}
M.~Alekseev {\em et~al.}, ``{Flavour Separation of Helicity Distributions from Deep Inelastic Muon-Deuteron Scattering},'' {\em Phys. Lett. B}, vol.~680, pp.~217--224, 2009.

\bibitem{COMPASS:2010hwr}
M.~G. Alekseev {\em et~al.}, ``{Quark helicity distributions from longitudinal spin asymmetries in muon-proton and muon-deuteron scattering},'' {\em Phys. Lett. B}, vol.~693, pp.~227--235, 2010.

\bibitem{STAR:2010xwx}
M.~M. Aggarwal {\em et~al.}, ``{Measurement of the parity-violating longitudinal single-spin asymmetry for $W^{\pm}$ boson production in polarized proton-proton collisions at $\sqrt{s} = 500-GeV$},'' {\em Phys. Rev. Lett.}, vol.~106, p.~062002, 2011.

\bibitem{STAR:2012hth}
L.~Adamczyk {\em et~al.}, ``{Longitudinal and transverse spin asymmetries for inclusive jet production at mid-rapidity in polarized $p+p$ collisions at $\sqrt{s}=200$ GeV},'' {\em Phys. Rev. D}, vol.~86, p.~032006, 2012.

\bibitem{STAR:2014afm}
L.~Adamczyk {\em et~al.}, ``{Measurement of longitudinal spin asymmetries for weak boson production in polarized proton-proton collisions at RHIC},'' {\em Phys. Rev. Lett.}, vol.~113, p.~072301, 2014.

\bibitem{STAR:2014wox}
L.~Adamczyk {\em et~al.}, ``{Precision Measurement of the Longitudinal Double-spin Asymmetry for Inclusive Jet Production in Polarized Proton Collisions at $\sqrt{s}=200$ GeV},'' {\em Phys. Rev. Lett.}, vol.~115, no.~9, p.~092002, 2015.

\bibitem{PHENIX:2010aru}
A.~Adare {\em et~al.}, ``{Event Structure and Double Helicity Asymmetry in Jet Production from Polarized $p+p$ Collisions at $\sqrt{s} = 200$\textasciitilde{}GeV},'' {\em Phys. Rev. D}, vol.~84, p.~012006, 2011.

\bibitem{PHENIX:2010rkr}
A.~Adare {\em et~al.}, ``{Cross Section and Parity Violating Spin Asymmetries of $W^\pm$ Boson Production in Polarized $p+p$ Collisions at $\sqrt{s}=500$ GeV},'' {\em Phys. Rev. Lett.}, vol.~106, p.~062001, 2011.

\bibitem{PHENIX:2015ade}
A.~Adare {\em et~al.}, ``{Measurement of parity-violating spin asymmetries in W$^{\pm}$ production at midrapidity in longitudinally polarized $p+p$ collisions},'' {\em Phys. Rev. D}, vol.~93, no.~5, p.~051103, 2016.

\bibitem{AbdulKhalek:2021gbh}
R.~Abdul~Khalek {\em et~al.}, ``{Science Requirements and Detector Concepts for the Electron-Ion Collider}: {EIC Yellow Report},'' {\em Nucl. Phys. A}, vol.~1026, p.~122447, 2022.

\bibitem{Achenbach:2023pba}
P.~Achenbach {\em et~al.}, ``{The Present and Future of QCD},'' 3 2023.

\bibitem{Abir:2023fpo}
R.~Abir {\em et~al.}, ``{The case for an EIC Theory Alliance: Theoretical Challenges of the EIC},'' 5 2023.

\bibitem{AbdulKhalek:2022hcn}
R.~Abdul~Khalek {\em et~al.}, ``{Snowmass 2021 White Paper: Electron Ion Collider for High Energy Physics},'' 3 2022.

\bibitem{Burkert:2022hjz}
V.~D. Burkert {\em et~al.}, ``{Precision studies of QCD in the low energy domain of the EIC},'' {\em Prog. Part. Nucl. Phys.}, vol.~131, p.~104032, 2023.

\bibitem{Anderle:2021dpv}
D.~P. Anderle, T.-J. Hou, H.~Xing, M.~Yan, C.~P. Yuan, and Y.~Zhao, ``{Determining the helicity structure of the nucleon at the Electron Ion Collider in China},'' {\em JHEP}, vol.~08, p.~034, 2021.

\bibitem{Quintans:2022utc}
C.~Quintans, ``{The New AMBER Experiment at the CERN SPS},'' {\em Few Body Syst.}, vol.~63, no.~4, p.~72, 2022.

\bibitem{deFlorian:2009vb}
D.~de~Florian, R.~Sassot, M.~Stratmann, and W.~Vogelsang, ``{Extraction of Spin-Dependent Parton Densities and Their Uncertainties},'' {\em Phys. Rev. D}, vol.~80, p.~034030, 2009.

\bibitem{Nocera:2014gqa}
E.~R. Nocera, R.~D. Ball, S.~Forte, G.~Ridolfi, and J.~Rojo, ``{A first unbiased global determination of polarized PDFs and their uncertainties},'' {\em Nucl. Phys. B}, vol.~887, pp.~276--308, 2014.

\bibitem{Ethier:2017zbq}
J.~J. Ethier, N.~Sato, and W.~Melnitchouk, ``{First simultaneous extraction of spin-dependent parton distributions and fragmentation functions from a global QCD analysis},'' {\em Phys. Rev. Lett.}, vol.~119, no.~13, p.~132001, 2017.

\bibitem{Ji:2013dva}
X.~Ji, ``{Parton Physics on a Euclidean Lattice},'' {\em Phys. Rev. Lett.}, vol.~110, p.~262002, 2013.

\bibitem{Ji:2014gla}
X.~Ji, ``{Parton Physics from Large-Momentum Effective Field Theory},'' {\em Sci. China Phys. Mech. Astron.}, vol.~57, pp.~1407--1412, 2014.

\bibitem{Ji:2020ect}
X.~Ji, Y.-S. Liu, Y.~Liu, J.-H. Zhang, and Y.~Zhao, ``{Large-momentum effective theory},'' {\em Rev. Mod. Phys.}, vol.~93, no.~3, p.~035005, 2021.

\bibitem{Fan:2022kcb}
Z.~Fan, W.~Good, and H.-W. Lin, ``{Gluon parton distribution of the nucleon from (2+1+1)-flavor lattice QCD in the physical-continuum limit},'' {\em Phys. Rev. D}, vol.~108, no.~1, p.~014508, 2023.

\bibitem{Chen:2016utp}
J.-W. Chen, S.~D. Cohen, X.~Ji, H.-W. Lin, and J.-H. Zhang, ``{Nucleon Helicity and Transversity Parton Distributions from Lattice QCD},'' {\em Nucl. Phys. B}, vol.~911, pp.~246--273, 2016.

\bibitem{Lin:2019ocg}
H.-W. Lin and R.~Zhang, ``{Lattice finite-volume dependence of the nucleon parton distributions},'' {\em Phys. Rev. D}, vol.~100, no.~7, p.~074502, 2019.

\bibitem{Lin:2018pvv}
H.-W. Lin, J.-W. Chen, X.~Ji, L.~Jin, R.~Li, Y.-S. Liu, Y.-B. Yang, J.-H. Zhang, and Y.~Zhao, ``{Proton Isovector Helicity Distribution on the Lattice at Physical Pion Mass},'' {\em Phys. Rev. Lett.}, vol.~121, no.~24, p.~242003, 2018.

\bibitem{Alexandrou:2020qtt}
C.~Alexandrou, K.~Cichy, M.~Constantinou, J.~R. Green, K.~Hadjiyiannakou, K.~Jansen, F.~Manigrasso, A.~Scapellato, and F.~Steffens, ``{Lattice continuum-limit study of nucleon quasi-PDFs},'' {\em Phys. Rev. D}, vol.~103, p.~094512, 2021.

\bibitem{Alexandrou:2016jqi}
C.~Alexandrou, K.~Cichy, M.~Constantinou, K.~Hadjiyiannakou, K.~Jansen, F.~Steffens, and C.~Wiese, ``{Updated Lattice Results for Parton Distributions},'' {\em Phys. Rev. D}, vol.~96, no.~1, p.~014513, 2017.

\bibitem{Alexandrou:2017huk}
C.~Alexandrou, K.~Cichy, M.~Constantinou, K.~Hadjiyiannakou, K.~Jansen, H.~Panagopoulos, and F.~Steffens, ``{A complete non-perturbative renormalization prescription for quasi-PDFs},'' {\em Nucl. Phys. B}, vol.~923, pp.~394--415, 2017.

\bibitem{Lin:2017ani}
H.-W. Lin, J.-W. Chen, T.~Ishikawa, and J.-H. Zhang, ``{Improved parton distribution functions at the physical pion mass},'' {\em Phys. Rev. D}, vol.~98, no.~5, p.~054504, 2018.

\bibitem{Alexandrou:2018pbm}
C.~Alexandrou, K.~Cichy, M.~Constantinou, K.~Jansen, A.~Scapellato, and F.~Steffens, ``{Light-Cone Parton Distribution Functions from Lattice QCD},'' {\em Phys. Rev. Lett.}, vol.~121, no.~11, p.~112001, 2018.

\bibitem{Fan:2020nzz}
Z.~Fan, X.~Gao, R.~Li, H.-W. Lin, N.~Karthik, S.~Mukherjee, P.~Petreczky, S.~Syritsyn, Y.-B. Yang, and R.~Zhang, ``{Isovector parton distribution functions of the proton on a superfine lattice},'' {\em Phys. Rev. D}, vol.~102, no.~7, p.~074504, 2020.

\bibitem{HadStruc:2022nay}
R.~G. Edwards {\em et~al.}, ``{Non-singlet quark helicity PDFs of the nucleon from pseudo-distributions},'' {\em JHEP}, vol.~03, p.~086, 2023.

\bibitem{Ji:2020brr}
X.~Ji, Y.~Liu, A.~Sch\"afer, W.~Wang, Y.-B. Yang, J.-H. Zhang, and Y.~Zhao, ``{A Hybrid Renormalization Scheme for Quasi Light-Front Correlations in Large-Momentum Effective Theory},'' {\em Nucl. Phys. B}, vol.~964, p.~115311, 2021.

\bibitem{LatticePartonCollaborationLPC:2021xdx}
Y.-K. Huo {\em et~al.}, ``{Self-renormalization of quasi-light-front correlators on the lattice},'' {\em Nucl. Phys. B}, vol.~969, p.~115443, 2021.

\bibitem{LatticeParton:2022xsd}
F.~Yao {\em et~al.}, ``{Nucleon Transversity Distribution in the Continuum and Physical Mass Limit from Lattice QCD},'' 8 2022.

\bibitem{Holligan:2023rex}
J.~Holligan, X.~Ji, H.-W. Lin, Y.~Su, and R.~Zhang, ``{Precision control in lattice calculation of x-dependent pion distribution amplitude},'' {\em Nucl. Phys. B}, vol.~993, p.~116282, 2023.

\bibitem{LatticeParton:2022zqc}
J.~Hua {\em et~al.}, ``{Pion and Kaon Distribution Amplitudes from Lattice QCD},'' {\em Phys. Rev. Lett.}, vol.~129, no.~13, p.~132001, 2022.

\bibitem{Su:2022fiu}
Y.~Su, J.~Holligan, X.~Ji, F.~Yao, J.-H. Zhang, and R.~Zhang, ``{Resumming quark's longitudinal momentum logarithms in LaMET expansion of lattice PDFs},'' {\em Nucl. Phys. B}, vol.~991, p.~116201, 2023.

\bibitem{Zhang:2023bxs}
R.~Zhang, J.~Holligan, X.~Ji, and Y.~Su, ``{Leading power accuracy in lattice calculations of parton distributions},'' {\em Phys. Lett. B}, vol.~844, p.~138081, 2023.

\bibitem{Zichichi:1979gj}
{\em {The Whys Of Subnuclear Physics. Proceedings Of The 1977 International School Of Subnuclear Physics, Held in Erice, Trapani, Sicily, July 23 - August 10, 1977}}, 1979.

\bibitem{Hasenfratz:2001hp}
A.~Hasenfratz and F.~Knechtli, ``{Flavor symmetry and the static potential with hypercubic blocking},'' {\em Phys. Rev. D}, vol.~64, p.~034504, 2001.

\bibitem{MILC:2012znn}
A.~Bazavov {\em et~al.}, ``{Lattice QCD Ensembles with Four Flavors of Highly Improved Staggered Quarks},'' {\em Phys. Rev. D}, vol.~87, no.~5, p.~054505, 2013.

\bibitem{Follana:2006rc}
E.~Follana, Q.~Mason, C.~Davies, K.~Hornbostel, G.~P. Lepage, J.~Shigemitsu, H.~Trottier, and K.~Wong, ``{Highly improved staggered quarks on the lattice, with applications to charm physics},'' {\em Phys. Rev. D}, vol.~75, p.~054502, 2007.

\bibitem{Gupta:2018qil}
R.~Gupta, Y.-C. Jang, B.~Yoon, H.-W. Lin, V.~Cirigliano, and T.~Bhattacharya, ``{Isovector Charges of the Nucleon from 2+1+1-flavor Lattice QCD},'' {\em Phys. Rev. D}, vol.~98, p.~034503, 2018.

\bibitem{Bhattacharya:2015wna}
T.~Bhattacharya, V.~Cirigliano, S.~Cohen, R.~Gupta, A.~Joseph, H.-W. Lin, and B.~Yoon, ``{Iso-vector and Iso-scalar Tensor Charges of the Nucleon from Lattice QCD},'' {\em Phys. Rev. D}, vol.~92, no.~9, p.~094511, 2015.

\bibitem{Bhattacharya:2015esa}
T.~Bhattacharya, V.~Cirigliano, R.~Gupta, H.-W. Lin, and B.~Yoon, ``{Neutron Electric Dipole Moment and Tensor Charges from Lattice QCD},'' {\em Phys. Rev. Lett.}, vol.~115, no.~21, p.~212002, 2015.

\bibitem{Bhattacharya:2013ehc}
T.~Bhattacharya, S.~D. Cohen, R.~Gupta, A.~Joseph, H.-W. Lin, and B.~Yoon, ``{Nucleon Charges and Electromagnetic Form Factors from 2+1+1-Flavor Lattice QCD},'' {\em Phys. Rev. D}, vol.~89, no.~9, p.~094502, 2014.

\bibitem{Bali:2016lva}
G.~S. Bali, B.~Lang, B.~U. Musch, and A.~Sch\"afer, ``{Novel quark smearing for hadrons with high momenta in lattice QCD},'' {\em Phys. Rev. D}, vol.~93, no.~9, p.~094515, 2016.

\bibitem{Babich:2010qb}
R.~Babich, J.~Brannick, R.~C. Brower, M.~A. Clark, T.~A. Manteuffel, S.~F. McCormick, J.~C. Osborn, and C.~Rebbi, ``{Adaptive multigrid algorithm for the lattice Wilson-Dirac operator},'' {\em Phys. Rev. Lett.}, vol.~105, p.~201602, 2010.

\bibitem{Osborn:2010mb}
J.~C. Osborn, R.~Babich, J.~Brannick, R.~C. Brower, M.~A. Clark, S.~D. Cohen, and C.~Rebbi, ``{Multigrid solver for clover fermions},'' {\em PoS}, vol.~LATTICE2010, p.~037, 2010.

\bibitem{Edwards:2004sx}
R.~G. Edwards and B.~Joo, ``{The Chroma software system for lattice QCD},'' {\em Nucl. Phys. B Proc. Suppl.}, vol.~140, p.~832, 2005.

\bibitem{Martinelli:1994ty}
G.~Martinelli, C.~Pittori, C.~T. Sachrajda, M.~Testa, and A.~Vladikas, ``{A General method for nonperturbative renormalization of lattice operators},'' {\em Nucl. Phys. B}, vol.~445, pp.~81--108, 1995.

\bibitem{Zhang:2020gaj}
R.~Zhang, C.~Honkala, H.-W. Lin, and J.-W. Chen, ``{Pion and kaon distribution amplitudes in the continuum limit},'' {\em Phys. Rev. D}, vol.~102, no.~9, p.~094519, 2020.

\bibitem{Holligan:2024umc}
J.~Holligan and H.-W. Lin, ``{Pion valence quark distribution at physical pion mass of N $_{f}$ = 2 + 1 + 1 lattice QCD},'' {\em J. Phys. G}, vol.~51, no.~6, p.~065101, 2024.

\bibitem{Chou:2022drv}
C.-Y. Chou and J.-W. Chen, ``{One-loop hybrid renormalization matching kernels for quasiparton distributions},'' {\em Phys. Rev. D}, vol.~106, no.~1, p.~014507, 2022.

\bibitem{Yao:2022vtp}
F.~Yao, Y.~Ji, and J.-H. Zhang, ``{Connecting Euclidean to light-cone correlations: from flavor nonsinglet in forward kinematics to flavor singlet in non-forward kinematics},'' {\em JHEP}, vol.~11, p.~021, 2023.

\bibitem{Izubuchi:2018srq}
T.~Izubuchi, X.~Ji, L.~Jin, I.~W. Stewart, and Y.~Zhao, ``{Factorization Theorem Relating Euclidean and Light-Cone Parton Distributions},'' {\em Phys. Rev. D}, vol.~98, no.~5, p.~056004, 2018.

\bibitem{Holligan:2023jqh}
J.~Holligan and H.-W. Lin, ``{Systematic Improvement of $x$-dependent Unpolarized Nucleon Generalized Parton Distribution in Lattice-QCD Calculation},'' 12 2023.

\bibitem{Gao:2021dbh}
X.~Gao, A.~D. Hanlon, S.~Mukherjee, P.~Petreczky, P.~Scior, S.~Syritsyn, and Y.~Zhao, ``{Lattice QCD Determination of the Bjorken-x Dependence of Parton Distribution Functions at Next-to-Next-to-Leading Order},'' {\em Phys. Rev. Lett.}, vol.~128, no.~14, p.~142003, 2022.

\bibitem{Chen:2018xof}
J.-W. Chen, L.~Jin, H.-W. Lin, Y.-S. Liu, Y.-B. Yang, J.-H. Zhang, and Y.~Zhao, ``{Lattice Calculation of Parton Distribution Function from LaMET at Physical Pion Mass with Large Nucleon Momentum},'' 3 2018.

\bibitem{Gao:2022uhg}
X.~Gao, A.~D. Hanlon, J.~Holligan, N.~Karthik, S.~Mukherjee, P.~Petreczky, S.~Syritsyn, and Y.~Zhao, ``{Unpolarized proton PDF at NNLO from lattice QCD with physical quark masses},'' {\em Phys. Rev. D}, vol.~107, no.~7, p.~074509, 2023.

\bibitem{Braun:2018brg}
V.~M. Braun, A.~Vladimirov, and J.-H. Zhang, ``{Power corrections and renormalons in parton quasidistributions},'' {\em Phys. Rev. D}, vol.~99, no.~1, p.~014013, 2019.

\bibitem{Moch:2004pa}
S.~Moch, J.~A.~M. Vermaseren, and A.~Vogt, ``{The Three loop splitting functions in QCD: The Nonsinglet case},'' {\em Nucl. Phys. B}, vol.~688, pp.~101--134, 2004.

\end{thebibliography}

\end{document}